%% file: main-prd.tex
\documentclass[aps,floatfix,prd,preprint]{revtex4-1}
\usepackage{graphicx,bm,amsmath,color,tabularx,amssymb}
\usepackage[utf8]{inputenc} 
\usepackage{comment}
\def\*{^{(*)}}
\def\be{\begin{equation}}
\def\ee{\end{equation}}
\def\dq{\frac{d^3q}{(2\pi)^3}\,}

\def\Lc{\Lambda_c}

\def\S{\Sigma_c}
\def\jp{J/\psi}
\def\jpp{\jp\, p}
\def\D{\bar D}

\def\>{\big>}
\def\<{\big<}
\def\|{\big\vert}

\def\etal{\textit{et~al.~}}

\begin{document}
\title{Production of $P_c$ states in $\Lambda_b$ decays}
\author{T.\,J.\,Burns}
\affiliation{Department of Physics, Swansea University, Singleton Park, Swansea, SA2 8PP, UK.}
\author{E.\,S.\,Swanson}
\affiliation{Department of Physics and Astronomy, University of Pittsburgh, Pittsburgh, PA 15260, USA.}

\begin{abstract}
We develop a model for the production of the $P_c$ states observed at  LHCb in $\Lambda_b\to\jpp\,K^-$ decays. With fewer parameters than other approaches, we obtain excellent fits to the $\jpp$ invariant mass spectrum, capturing both the prominent peaks, and broader features over the full range of invariant mass. A distinguishing feature of our model is that whereas $P_c(4312)$, $P_c(4380)$ and $P_c(4440)$ are resonances with $\S\*\D\*$ constituents, the nature of $P_c(4457)$ is quite different, and can be understood either as a $\S\D^*$ threshold cusp, a $\Lc(2595)\D$ enhancement due to the triangle singularity, or a  $\Lc(2595)\D$ resonance. We propose experimental measurements that can discriminate among these possibilities. Unlike in other models, our production mechanism respects isospin symmetry and the empirical dominance of colour-enhanced processes in weak decays, and additionally gives a natural explanation for the overall shape of the data. Our model is consistent with experimental constraints from  photoproduction and $\Lambda_b\to \Lc\D^{(*)0}K^-$ decays and it does not imply the existence of partner states whose apparent absence in experiments is unexplained in other models. \end{abstract}

\maketitle 

\section{Introduction}
\label{Sec:introduction}

Much of the considerable literature on the LHCb $P_c$ states relates to their mass spectrum, quantum numbers, and decays. In this paper we develop a model which includes all of these features, but also goes further, in aiming to fit the $\jpp$ mass spectrum in $\Lambda_b\to J/\psi\, p \,K^-$. In this sense our remit is similar to that of refs \cite{Du:2019pij,Du:2021fmf,Nakamura:2021dix,Nakamura:2021qvy,Kuang:2020bnk}, but with some notable differences which we describe throughout the paper. 

The $P_c$ states have been widely interpreted as molecular states with $\S\*\D\*$ constituents. However we have recently argued that models that describe all of the $P_c$ states exclusively in terms of $\S\*\D\*$ degrees of freedom suffer from various phenomenological problems when confronted with experimental data~\cite{Burns:2021jlu}. In this paper we develop a model which resolves these problems, and gives an excellent fit to the experimental data on $\Lambda_b\to J/\psi\, p \,K^-$ decays. 

One of the issues with the commonly held view is that the direct production of $\S\*\D\*$ constituents in $\Lambda_b$ decays requires either isospin violation, or a colour-suppressed weak transition. A natural resolution of this problem, which is the foundation of our model, is to produce the $P_c$ states via channels with $\Lc\D$ flavour, such as $\Lc\D$, $\Lc\D^*$, and $\Lc(2595)\D$. In this case the production conserves isospin, and is colour-enhanced. In our previous paper, we concluded that experimental results imply that the $P_c$ states decay dominantly to $\Lc\D^*$, which gives further support to this production mechanism.

There are in addition several other reasons to expect an important role for channels with $\Lc\*\D\*$ flavour. The presumed importance of $\S\*\D\*$ constituents is usually attributed to the proximity of the $P_c$ masses to the corresponding thresholds. But in the case of $P_c(4457)$, the proximity to $\Lc(2595)\D$ threshold is even more striking, with a difference of just 0.2~MeV in the central values. This is a strong indication for the possible role of $\Lc(2595)\D$ degrees of freedom, an idea which we and others have explored in previous work \cite{Burns:2015dwa,Geng:2017hxc,Burns:2019iih,Yalikun:2021bfm}.

Another distinguishing feature of our approach is that, in contrast to many other models, we do not assume that $P_c(4457)$ is a $\S\D^*$ bound state. Here we are mainly being guided by experimental data: whereas $P_c(4440)$ is unambiguously below $\S^+\D^{*0}$ threshold, the same is not true of $P_c(4457)$. Moreover, we argued \cite{Burns:2021jlu} that a model in which both $P_c(4440)$ and $P_c(4457)$ are bound $\S\D^*$ states is difficult to reconcile with experimental data, because it implies that one of the states would decay prominently to $\Lc\D$ (contradicting experimental data), and implies there should be partner states near the $\S^*\D^*$ threshold (also absent from data). We showed that these problems can be avoided by assuming that only $P_c(4440)$ is a $\S\D^*$ bound state, in which case an alternative explanation for $P_c(4457)$ is needed. Because its mass is consistent with both the $\S^+\D^{*0}$ and $\Lc(2595)\D$ thresholds, there are several viable alternative scenarios: it could be a threshold cusp arising from $\S\D^*\to\jpp$ or $\Lc(2595)\D\to \jpp$, an enhancement due to a triangle singularity in the $\Lc(2595)\D$ loop diagram or, as in our earlier paper \cite{Burns:2019iih}, a resonance with $\Lc(2595)\D$ degrees of freedom. We will explore all of these alternative scenarios for $P_c(4457)$, and ultimately find excellent agreement with $\Lambda_b\to J/\psi\, p \,K^-$ data.

In Section~\ref{sec:model} we introduce our model, starting with a description of the production mechanism, and then describing how the generic features of the model can describe the $\Lambda_b\to J/\psi\, p \,K^-$ data, while also satisfying other experimental constraints which very likely eliminate broad categories of previous modelling. We then give details of the amplitudes and their calculation. Results of model fits for five different cases are presented in Section~\ref{sec:results}. These cases systematically add complexity so that model accuracy can be evaluated against model efficiency. Finally in Section~\ref{sec:conclusions} we conclude and give suggestions for future experimental study of $P_c$ states.

\section{Model}
\label{sec:model}

\subsection{Production}
\label{sec:production}

\begin{figure}[ht]
    \includegraphics[width=\textwidth]{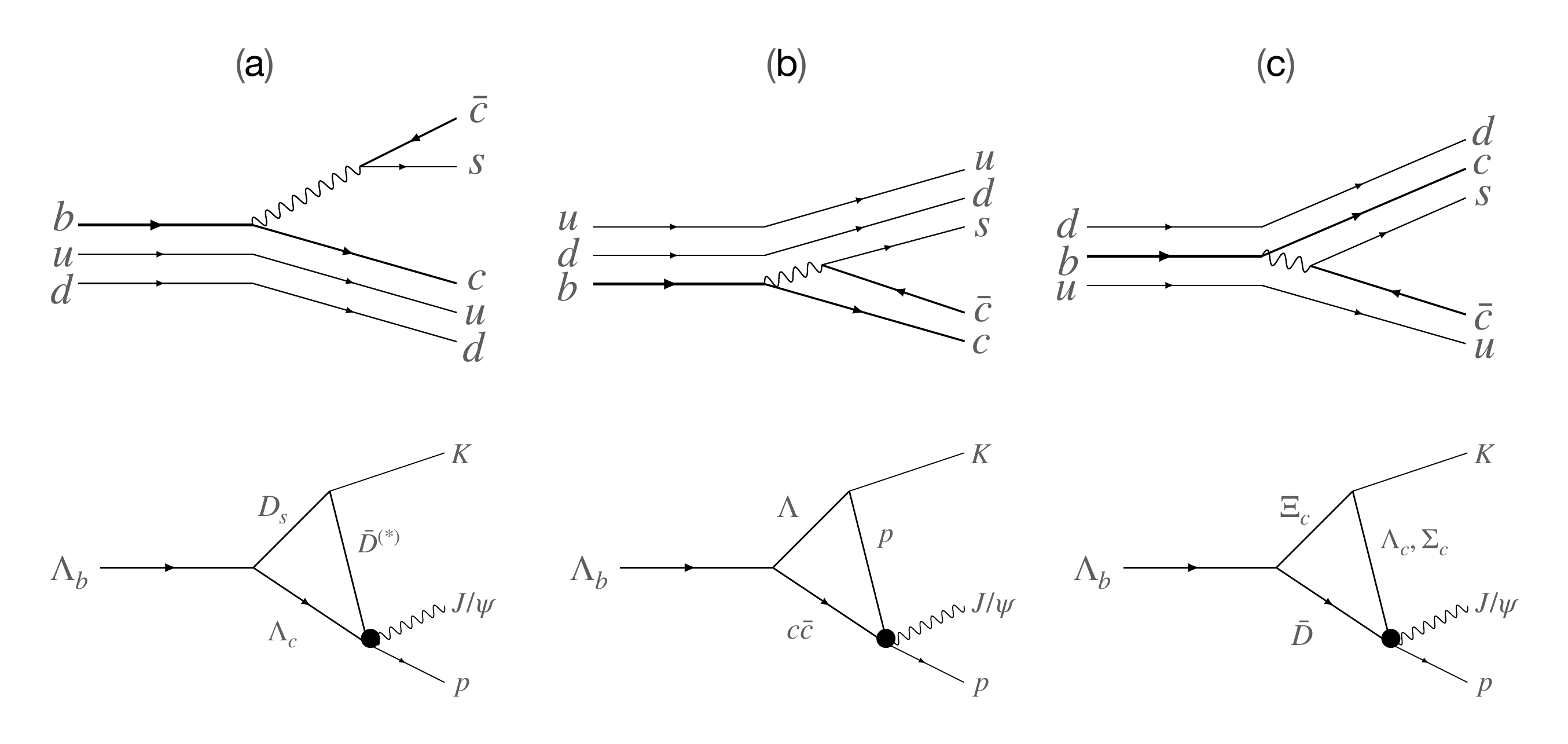}
    \flushleft $\Lambda_b$ vertex:\qquad\qquad Large\qquad\qquad\qquad\qquad\qquad\qquad Small\qquad\qquad\qquad\qquad\qquad Small\\
    $P_c$ vertex:\qquad\qquad Large\qquad\qquad\qquad\qquad\qquad\qquad Small\qquad\qquad\qquad\qquad\qquad Large
   \caption{Production mechanisms. The top panel shows the quark line diagrams at the electroweak vertex, and the bottom panels show the corresponding production diagram for $P_c$ states.}
    \label{fig:production}
\end{figure}

In our previous paper~\cite{Burns:2019iih} we noted that there are three possible quark line topologies for the $\Lambda_b$ weak vertex, and each has an associated loop diagram that could contribute to the $\jpp$ spectrum. The possibilities are shown in Figure~\ref{fig:production}, where the labels describe flavour only; hence ``$\Lc$'' can stand for $\Lc$, $\Lc(2595)$ or $\Lc(2625)$, for example. For simplicity, in our diagrams and much of the discussion, we will ignore charge labels on states; so for example ``$K$'' means $K^-$, and ``$D_s$'' means $D_s^-$.

Below each diagram we have classified the $\Lambda_b$ vertex (the weak decay) and the $P_c$ vertex (generating $\jpp$) as ``large'' or ``small'', based on simple arguments that we outline below. On the basis of these classifications, we will assume that production proceeds via diagram~(a), being the only diagram for which both vertices are ``large''.

Our classifications for the $\Lambda_b$ vertex follow the arguments in our previous paper~\cite{Burns:2019iih}: diagram (a) is colour-favoured, and hence is enhanced with respect to diagrams (b) and (c), for which the weak vertices are colour-suppressed.

There are strong indications (experimental and theoretical) supporting the dominance of colour-favoured diagrams. Indeed the largest measured two-body decay of $\Lambda_b$ is $\Lc^+D_s^-$, corresponding to diagram (a), with branching fraction $1.1\pm 1.0\%$. Branching fractions for decays such as $\Lambda\jp$, corresponding to diagram (b), have not been measured directly, but ref.~\cite{Zhu:2018jet} finds $\mathcal B(\Lambda_b\to \Lambda\jp)=(3.72\pm 1.07)\times 10^{-4}$, using the measured product branching fraction  $\mathcal B(\Lambda_b\to \Lambda\jp)\times \mathcal B (b\to \Lambda_b)$ and the Heavy Flavour Averaging group value \cite{HFLAV:2016hnz} for the production rate $\mathcal B (b\to \Lambda_b)$. The result, which is comparable to theory predictions~\cite{Cheng:1996cs,Hsiao:2015txa,Hsiao:2015cda,Gutsche:2018utw,Zhu:2018jet}, indicates significant colour suppression of diagram (b) compared to diagram (a). Decays such as $\Lambda_b\to \Xi_c\D^*$, corresponding to diagram (c), have also not been measured directly, but naively we may expect these to be comparable in magnitude to diagram (b), and indeed this is what we found in a quark model calculation \cite{Burns:2019iih}.

Du~\etal\cite{Du:2021fmf} have questioned the scale of colour suppression in $\Lambda_b$ decays, on the basis of a comparison with $\Lambda_c$ decays. We notice, however, that direct comparisons between $\Lambda_c$ and $\Lambda_b$ decays are not reliable.  For example, the phenomenological analyses of refs.~\cite{Zhao:2018mov,Hsiao:2021nsc} suggest that in $\Lambda_c$ decays, the analogues of diagrams (a), (b) and (c) are comparable in magnitude, which is very different to the situation in $\Lambda_b$ decays, where there is empirical evidence (noted above) for significant suppression of diagram (b) compared to (a). 

Moreover, $\Lambda_c$ and $\Lambda_b$ decays involve different decay diagrams, which makes direct comparison of similar modes impossible. The argument of Du~\etal is that colour suppression would imply 
that $\Lambda_c^+\to \Sigma^0 K^+$ is suppressed compared to  $\Lambda_c^+\to \Lambda K^+$, as the former is produced in the analogue of diagram (c) but not (a), and they notice that this is not consistent with data. The analogy does not work, however, because in the $\Lambda_c$ decays there are additional $W$-exchange diagrams that are absent from the corresponding $\Lambda_b$ decays, and their contribution is known to be significant, as evidenced (for example) by the abundance of modes such as  $\Xi^0K^+$ and $\Delta^{++}K^-$, which are produced only via $W$-exchange diagrams~\cite{Kohara:1991ug,Chau:1995gk,Cheng:1996cs,Cheng:2018hwl}.

Having argued that analogies with related systems can be misleading, it is nonetheless interesting to compare $b\to c s\bar c$ transitions in baryon and meson systems, because in the latter case the significant magnitude of colour suppression is abundantly clear in the experimental data. See, for example, the experimental data summarised in Fig.~3 in Ref.~\cite{Burns:2020xne}, illustrating that  colour-favoured modes ($B\to D_s\*\D\*$) are enhanced by one or two orders of magnitude compared to colour-suppressed modes ($B\to c\bar c\, K\*$).

Additionally, there is indirect evidence for colour suppression in $\Lambda_b$ decays from the measured branching fractions in $\Lambda_b\to \Lc\D^{(*)0}K^-$ decays~\cite{Stahl:2018eme}. As shown in Fig.~\ref{fig:production}, whereas the two-body decay $\Lambda_b^0\to\Lc^+D_s^-$ is a colour-favoured diagram of type (a), the three-body decays $\Lambda_b\to \Lc\D^{(*)0}K^-$ can occur via both of diagrams (a) and (c). (Notice that the $ \Lc\D^{(*)0}K^-$ combination appears as an intermediate state in the loop diagrams (a) or (c).) On the other hand, for the analogous meson decays, where $ud$ is replaced with $\bar d$, both the two-body mode $B^0\to D^+ D_s^-$, and the three-body modes $B^0\to D^+\D^{(*)0}K^-$, can only occur via the diagram~(a)---there is no analogue of diagram (c). Accordingly, a comparison of the ratios
\begin{align}
    \mathcal R\*_{\Lambda_b^0}&=\frac{\mathcal B(\Lambda_b^0\to \Lc^+\D^{(*)0}K^-)}{\mathcal B(\Lambda_b^0\to \Lc^+D_s^-)}\textrm{, and}\\
    \mathcal R\*_{\bar B^0}&=\frac{\mathcal B(\bar B^0\to D^+\D^{(*)0}K^-)}{\mathcal B(\bar B^0\to D^+D_s^-)}
\end{align}
can give an indication of the importance of diagram (c). Strikingly, the experimental ratios are found to be consistent~\cite{Stahl:2018eme}, which supports the hypothesis that diagram (c) is sub-dominant with respect to diagram (a).

We now turn to the second aspect of the classification of diagrams, namely the coupling of the triangle diagrams to the $P_c$ states. From the analysis in our previous paper~\cite{Burns:2021jlu}, $P_c(4312)$ decays overwhelming to $\Lc\D^*$, and hardly at all to $\jpp$. Heavy quark symmetry then implies a similar pattern for all of the $P_c$ states composed of $\S\*\D\*$ degrees of freedom, namely they couple much more strongly to $\Lc\D$ and $\Lc\D\*$ than to closed-charm channels such as $\jpp$ and $\eta_c\, p$. (Note that $P_c(4312)$ is a special case, which couples strongly to $\Lc\D^*$ but not $\Lc\D$, due to a selection rule~\cite{Voloshin:2019aut,Burns:2021jlu}.) On this basis we conclude that the $P_c$ vertices with $\Lc\D\*$ states in the triangle (diagrams (a) and (c)) are ``large'', whereas those with closed-charm (diagram (b)) are ``small''. 

Of course in diagram (c) there are contributions not only from $\Lc\D\*$, but also $\S\*\D\*$. Assuming the $P_c$ states are dominated by $\S\*\D\*$ degrees of freedom, then if $\S\*\D\*\to\S\*\D\*$ couplings are much stronger than $\Lc\D\*\to\S\*\D\*$, then conceivably the production of $P_c$ states could be dominated by diagram (c), despite the smaller $\Lambda_b$ vertex. However on general grounds we expect that the $\S\*\D\*\to\S\*\D\*$ and $\Lc\D\*\to\S\*\D\*$ couplings are comparable. 
For example, the $P_c(4312)$ width is due to $\S\D\to\Lc\D^*$, whereas the $P_c(4440)$ width, which is around twice as large, is due to both $\S\D^*\to\Lc\D^*$  and $\S\D^*\to \S^*\D$. Very roughly this suggests that  $\S\*\D\*\to\Lc\D\*$ and $\S\*\D\*\to\S\*\D\*$ have comparable magnitude. Indeed in the quark model, there is a direct relation between the one-pion exchange potentials for the corresponding potentials where $\Lc$ is replaced with $\S$, consistent with the expectation that their magnitudes are comparable:
\begin{equation}
\<\S\*\D\*\|V\|\Lc\D\*\>=-\frac{3}{4}\<\S\*\D\*\|V\|\S\D\*\>.
\end{equation}
Further comments on this point will be made in Section \ref{sect:disc}. 

For diagrams involving related channels such as $\Lc(2595)\D\*$ and $\Lc(2625)\D\*$, we note the enhancement of the rate where the $\jpp$ invariant mass coincides with the corresponding threshold, due to the small energy denominators . This is particularly striking in the case of $P_c(4457)$ which, as noted previously, overlaps with the $\Lc(2595)\D$ threshold.

\subsection{Overview}
\label{sec:overview}

\begin{figure}[ht]
    \centering
    \includegraphics[width=0.32\textwidth]{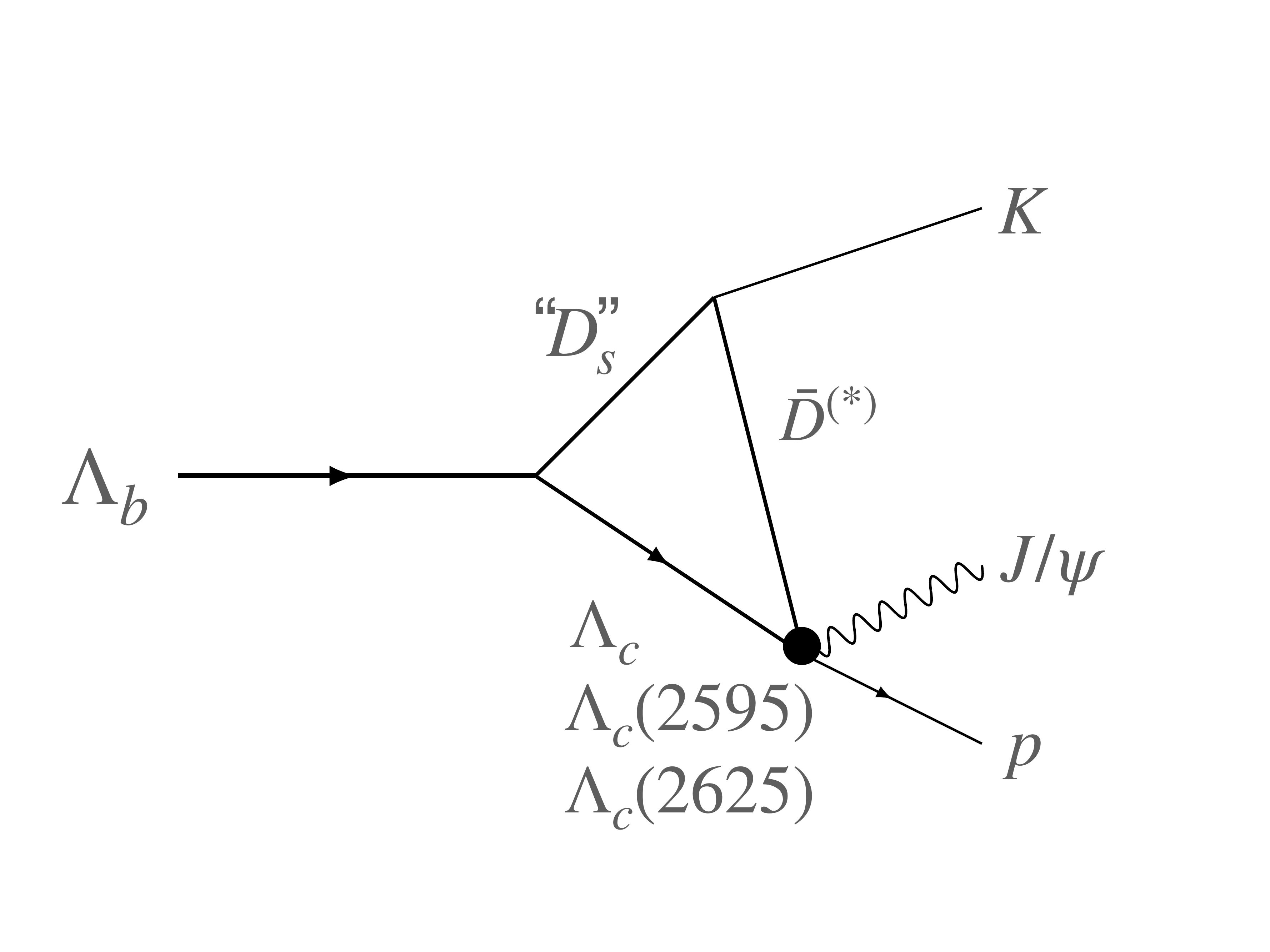} \qquad
    \includegraphics[width=.6\textwidth]{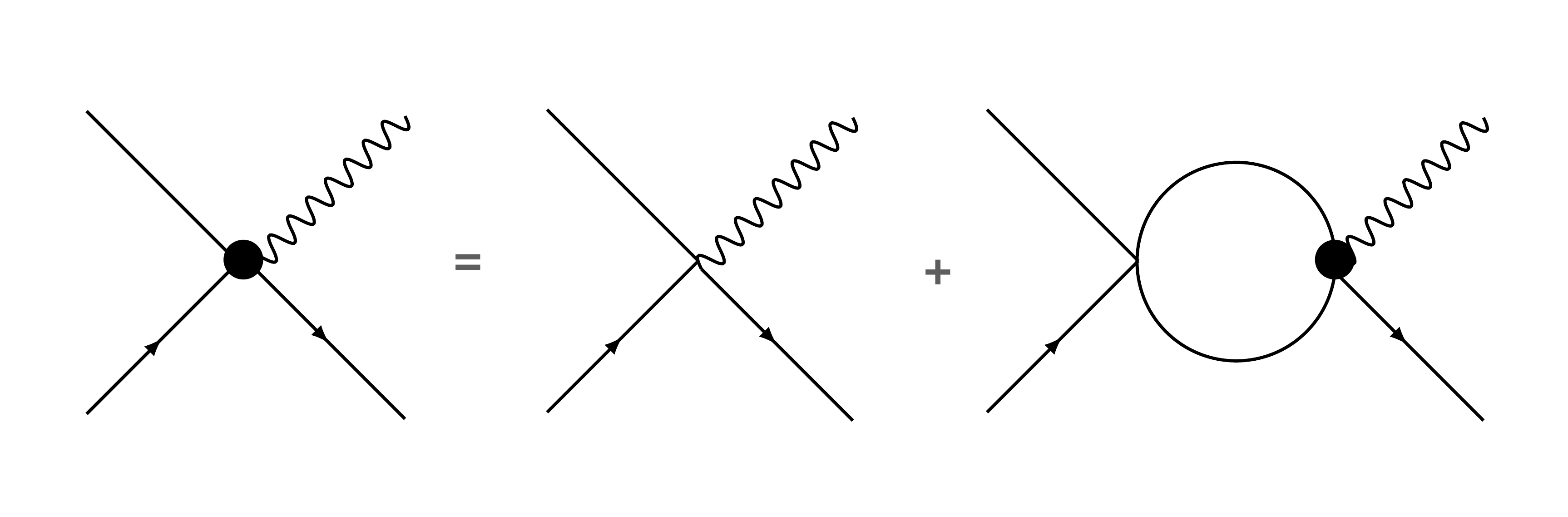}
    \caption{Production of $P_c$ states in $\Lambda_b$ decay through a combination of (left) a triangle diagram with a colour-favoured weak transition and (right) non-perturbative final state interactions.}
    \label{fig:prod8}
\end{figure}

Our model for the production of $P_c$ states in $\Lambda_b$ decay is depicted in Figure~\ref{fig:prod8}. Following the arguments in the previous section, we assume production via diagram (a), so that the triangle diagram features combinations with flavour $\Lc\*\D\*$, not $\S\*\D\*$ (Figure~\ref{fig:prod8}(left)). These intermediate states couple to the $\jpp$ final state either directly, or via non-perturbative final state interactions, which we implement via iterated bubble diagrams (Figure~\ref{fig:prod8}(right)). The $\S\*\D\*$ channels, which are important in explaining the $P_c$ states, enter via the bubble diagrams, for example via the $\Lc\*\D\*\to\S\*\D\*$ coupling.

Below we discuss in more detail the ingredients in the calculation, including the choice of hadrons to include in triangle and bubble diagrams, the nature of the ``$D_s$'' meson, and the model used for the interaction vertices. First, we make some general observations on how the model described here offers the possibility of describing not only the $P_c$ peaks, but also also other features in the  $\Lambda_b\to\jpp\, K^-$ data.

\begin{figure}[ht]
    \centering
    \includegraphics[width=\textwidth]{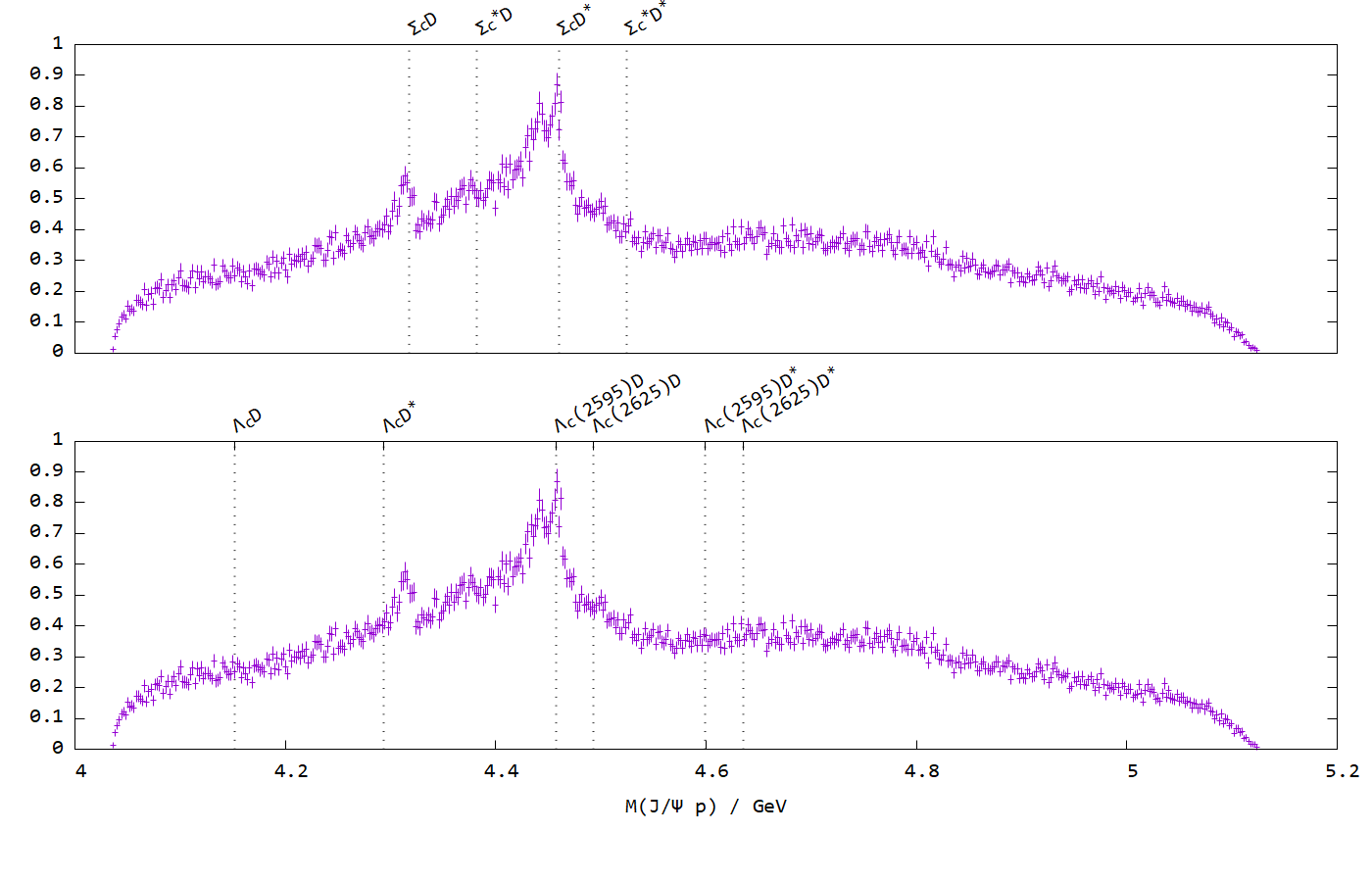}
\caption{The ``cos-weighted'' $\Lambda_b\to J/\psi\, p \,K^-$ spectrum from LHCb~\cite{Aaij:2019vzc}, overlaid with the thresholds for $\S\*\D\*$ (upper panel) and $\Lc\*\D\*$ (lower panel) channels.}
    \label{fig:thresholds}
\end{figure}

The experimental data is shown in Figure~\ref{fig:thresholds}, overlaid with the thresholds for channels with flavour $\S\*\D\*$ (top panel) and $\Lc\*\D\*$ (bottom panel). The sharp features corresponding to the $P_c(4312)$, $P_c(4440)$, and $P_c(4457)$ are evident, as well as the broader feature previously identified as  $P_c(4380)$. Models for $P_c$ states have concentrated overwhelming on the role of  $\S\*\D\*$ channels, based on the proximity of $P_c(4312)$,  $P_c(4380)$, $P_c(4440/4457)$ to $\S\D$, $\S^*\D$,  and $\S\D^*$ thresholds, respectively. We have argued on theoretical grounds for the importance of $\Lc\*\D\*$ channels, and we also notice that by including these channels, the model has leverage over a wider range of $\jpp$ invariant mass (comparing the top and bottom panels of Fig.~\ref{fig:thresholds}).

Having $\Lc\*\D\*$ channels in the triangle diagram also brings new possibilities for explaining not only the $P_c$ peaks, but also other features in the data. In our model, the leading order (triangle) diagram generates a cusp at the thresholds of the corresponding intermediate states with $\Lc\D$ flavour; this diagram may account for the shoulders in the data around the thresholds for $\Lc\D$, $\Lc(2595)\D^*$, or  $\Lc(2625)\D^*$, or indeed for the $P_c(4457)$ peak near $\Lc(2595)\D$. Another possibility, of particular relevance to $P_c(4457)$, is the logarithmic triangle singularity which, given a suitably chosen ``$D_s$'' mass, generates a strong enhancement at (or above) $\Lc(2595)\D$ threshold. All of these possibilities depend critically on the assumed production mechanism, involving $\Lc\*\D\*$ degrees of freedom.

Features associated with $\S\*\D\*$ degrees of freedom  appear via the bubble diagrams. Such diagrams generate cusps at the thresholds for the states in the bubbles, and we may expect in particular an important role here for the $P_c(4457)$ peak near $\S\D^*$ threshold. The non-perturbative sum over bubble diagrams can also generate resonance poles, and we expect these to be particularly important in accounting for structures further from threshold, namely $P_c(4312)$, $P_c(4380)$ and $P_c(4440)$. Similarly we also consider the possibility of a resonance with $\Lc(2595)\D$ degrees of freedom, which would arise from iteration of $\Lc(2595)\D\to \Lc(2595)\D$ bubble diagrams.

Before elaborating further on our model, it is worth highlighting how our (well-motivated) assumptions compare to those of other models in the literature. Our production mechanism conserves isospin and is colour-enhanced. By contrast, the model of Du \textit{et al.}~\cite{Du:2021fmf,Du:2019pij} assumes a point-like vertex, 
$
\langle \Lambda_B | H_{EW} | K \S\*\D\*\rangle
$, which requires either isospin breaking, or a colour-suppressed mechanism. With reference to Fig.~\ref{fig:production}, such a vertex could be obtained from diagram (a), but with an infinitely heavy $D_s$ and with  $\Lc$ replaced by $\Sigma_c$ (violating isospin), or via diagram (c), with an infinitely heavy $\Xi_c$ (and as noted, this diagram is colour-suppressed). 

The model of Nakamura~\cite{Nakamura:2021qvy}, like ours, assumes the dominance of colour-favoured processes. However this is implemented via a four-point function, analogous to diagram (a) with an infinite mass $D_s$.  This vertex is incorporated in a two-loop ``double triangle" diagram to generate the $\jpp\, K^-$ final state. Saturating the four-point vertex with $D_s$ exchange brings this rather elaborate mechanism into closer contact with our model. However the virtual pion present in the two-loop diagram represents final state interactions in the $\Sigma_c  K - \Lambda_c K$ system, in contrast with our final state interactions in the $\Sigma_c^{(*)} \bar{D}^{(*)} - \Lambda_c^{(*)} \bar{D}^{(*)}$ system, which is of course more natural for describing $\Sigma_c^{(*)}\D^{(*)}$ bound states with important couplings to $\Lambda_c^{(*)}\D^{(*)}$.

\subsection{Phenomenology}
\label{sec:phenomenology}

Although our primary focus in this paper is the  $\Lambda_b\to J/\psi\, p \,K^-$ data, there are additional experimental constraints which need to be considered in constructing a viable model for $P_c$ states. In particular, we have recently shown that experimental measurements on photoproduction, and $\Lambda_b$ decays to $ \Lambda_c \bar{D}^{(*)0} K^-$ and $\eta_c\, p\, K^-$, place severe constraints on putative models for $P_c$ states~\cite{Burns:2021jlu}. 

\begin{table}
    \centering
    \begin{tabularx}{\textwidth}{XXXXX}
\hline
&$P_c(4312)$&$P_c(4380)$&$P_c(4440)$&$P_c(4457)$\\
\hline
Scenario~A & $1/2^-~\S\D$&$3/2^-~\S^*\D$& $1/2^-~\S\D^*$& $3/2^-~\S\D^*$\\
Scenario~B & $1/2^-~\S\D$&$3/2^-~\S^*\D$& $3/2^-~\S\D^*$& $1/2^-~\S\D^*$\\
Scenario~C & $1/2^-~\S\D$&$3/2^-~\S^*\D$& $3/2^-~\S\D^*$& varies\\
\hline
    \end{tabularx}
    \caption{Quantum numbers and degrees of freedom in  various scenarios.}
    \label{tab:scenarios}
\end{table}

In our discussion we referred to three scenarios  involving binding in $\Sigma_c^{(*)}\bar{D}^{(*)}$ systems, as shown in Table \ref{tab:scenarios}~\cite{Burns:2021jlu}.
In Scenarios~A and B, which are common in the literature, all of the $P_c$ states are associated with attractive $\S\*\D\*$ interactions, and in particular both $P_c(4440)$ and $P_c(4457)$ are assumed to be (dominantly) $\S\D^*$ bound states, differing in the assignment of $1/2^-$ and $3/2^-$ quantum numbers. In Scenario~C, by contrast, we no longer assume that $P_c(4457)$ is a $\S\D^*$ bound state. This is partly inspired by the experimental reality: whereas $P_c(4440)$ is manifestly bound with respect to $\S\D^*$ threshold, $P_c(4457)$ is not. Indeed its mass is consistent with both $\S\D^*$ and $\Lc(2595)\D$ thresholds, and this implies several possible interpretations, all of which arise naturally with our proposed production mechanism: it could be a $\S\D^*$ or $\Lc(2595)\D$ cusp, a $\Lambda_c(2595)\bar{D}$ resonance, or a $\Lambda_c(2595)\bar{D}$ enhancement due to the logarithmic triangle singularity.

As well as being very natural, the alternatives incorporated in Scenario~C avoid serious phenomenological problems with Scenarios~A and B~\cite{Burns:2021jlu}. A key observation is the striking disparity between the $P_c$ branching fractions for $\Lc\D$ (of order 1\% or less) and $\Lc\D^*$ (as much as 59\%-87\% in the case of $P_c(4312)$). The suppression of $\Lc\D$ has a natural explanation in the case of $P_c(4312)$ (where it is due to a selection rule \cite{Voloshin:2019aut}), and for states with $3/2^-$ quantum numbers (where it is due to D-wave suppression). But for the $1/2^-$ $\S\D^*$ states of Scenarios~A and B, the mismatch between $\Lc\D$ and $\Lc\D^*$ is strikingly inconsistent with heavy quark symmetry, according to which the modes should be comparable. On this basis we exclude Scenarios~A and B, and are led to Scenario~C, which avoids this problem by not having a $1/2^-$ $\S\D^*$ bound state.

Another issue with Scenarios~A and B is that if the potentials are tuned to generate $\S\D^*$ bound states in both $1/2^-$ and $3/2^-$ channels, heavy quark symmetry implies there should also be $\S^*\D^*$ bound states with $1/2^-$, $3/2^-$, and $5/2^-$ quantum numbers, and such states are conspicuously absent from the experimental data (Fig.~\ref{fig:thresholds}). The problem does not arise in Scenario~C, for which the only $\S^*\D^*$ state that binds has $5/2^-$ quantum numbers, and its absence can be explained by the  suppressed D-wave decay (and production, as we argue later).

In all the scenarios, there is a $3/2^-$ $\S^*\D$ state, and this will be associated with the broad structure identified as $P_c(4380)$ in the original LHCb analysis \cite{Aaij:2015tga}. Note that, as explained in the appendix of ref.~\cite{Aaij:2019vzc}, the measured properties of $P_c(4380)$ are now regarded as obsolete, hence we concentrate on reproducing the data, rather than the measured mass and width. From a model perspective, a $3/2^-$ $\S^*\D$ state is an inevitable consequence of heavy-quark symmetry, since the diagonal potential in this channel is identical to that of $1/2^-$ $\S\D$, which is necessarily bound to account for the $P_c(4312)$.

Among the models which, like ours, aim to fit the $\Lambda_b\to\jpp\, K^-$ spectra, the model of
Du \textit{et al.}~\cite{Du:2021fmf,Du:2019pij} is one in which the states are classified according to Scenarios~A and B, and so has the associated difficulties which we outlined above, and which are explained in more detail in ref.~\cite{Burns:2021jlu}. The models of Kuang~\etal\cite{Kuang:2020bnk} and Nakamura~\etal\cite{Nakamura:2021qvy,Nakamura:2021dix} do not assume a common structure for $P_c(4312)$ and $P_c(4440)$, and in this sense they avoid the problems associated with Scenarios~A and B, but at the cost of ignoring (or contradicting) heavy quark symmetry, according to which molecular states do not appear in isolation, but as part of multiplets.

\subsection{Diagrams}
\label{sec:diagrams}

With the production mechanism in place, we are prepared to elaborate the amplitude model. We will assume that the $P_c$ states are produced via diagram (a), namely through hadron combinations of ``$\Lc\D$'' flavour. In particular, we will consider the ground state combinations $\Lc\D$ and $\Lc\D^*$, which are presumably most prominent on the production side (the weak and strong vertices), and also because at the non-perturbative interaction vertex they couple in S-wave to the assumed $1/2^-$ and $3/2^-$ quantum numbers of $P_c(4312/4380/4440)$. Since these channels do not couple to $5/2^-$ in S-wave, we are not including the $5/2^-$ channel in our analysis.

For reasons already discussed, we should also consider production via $\Lc(2595)\D$. Assuming S-wave dominance, this is a $1/2^+$ channel, which naturally suggests a possible role for the related channels $\Lc(2595)\D^*$ and $\Lc(2625)\D\*$, which can also be produced via the colour-favoured mechanism. Of these, we will consider only the $1/2^+$ combinations $\Lc(2595)\D^*$ and $\Lc(2625)\D^*$, since these appear in the same non-perturbative interaction matrix as the previous channel $\Lc(2595)\D$, and because, as noted previously, these degrees of freedom may also help to capture the features in the data near the corresponding thresholds (see Fig.~\ref{fig:thresholds}). To avoid overly complicating our model, we do not consider the related combinations with $3/2^+$ and $5/2^+$ quantum numbers.

We will explore a number of different cases of increasing complexity. The simplest cases involve only $1/2^-$ and $3/2^-$ channels, and we then add the $1/2^+$ channels. 

We will construct the amplitudes in the partial wave basis, and assume the minimal possible orbital angular momentum for the final state. For $1/2^-$ and $3/2^-$ the $\jpp$ state is $^2S_{1/2}$ and $^4S_{3/2}$, respectively, whereas for $1/2^+$ there are two possibilities, $^2P_{1/2}$ and $^4P_{1/2}$.

\begin{figure}
    \centering
   \includegraphics[width=\textwidth]{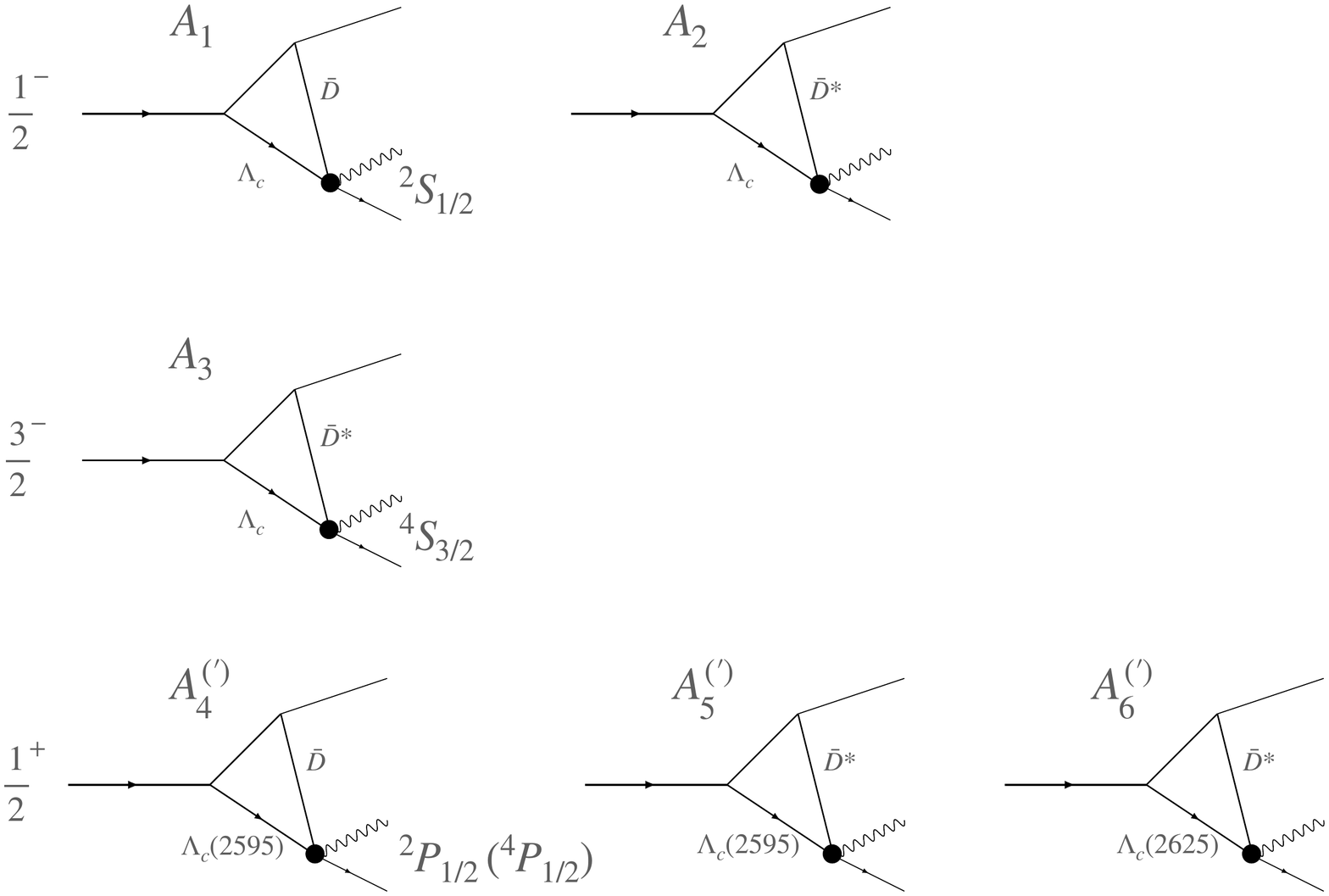}
    \caption{Amplitudes contributing to the $\Lambda_b\to \jpp\, K^-$ spectrum. The $1/2^-$ and $3/2^-$ channels are produced through $\Lc\D\*$ with $\jpp$ final states in $^2S_{1/2}$ (amplitudes $A_1, A_2$) or $^4S_{3/2}$ (amplitude $A_3$). The $1/2^+$ channels have production through $\Lc(2595)\D\*$ or $\Lc(2625)\D^*$ with $\jpp$ in $^2P_{1/2}$~(amplitudes $A_4,A_5,A_6$) or $^4P_{1/2}$~(amplitudes $A_4^\prime,A_5^\prime,A_6^\prime$). The unlabelled line refers to a $D_s^-$ meson, with interpretation described in Sect.\ref{sec:amplitudes}.}
    \label{fig:amps}
\end{figure}

With these assumptions, the final amplitudes that define our model are shown in  Fig.~\ref{fig:amps}. In $1/2^-$ there are two production channels ($\Lc\D$ and $\Lc\D^*$), and  two corresponding amplitudes (labelled $A_1$ and $A_2$), while in $3/2^-$ there is a single production channel  ($\Lc\D^*$) and corresponding amplitude ($A_3$). For $1/2^+$ there are three production channels ($\Lc(2595)\D,\Lc(2595)\D^*,\Lc(2625)\D^*$), hence three amplitudes ($A_4,A_5,A_6$) for the $^2P_{1/2}$ final state, and a further three ($A_4^\prime,A_5^\prime,A_6^\prime$) for
$^4P_{1/2}$. 

In each diagram, the filled vertex describes a non-perturbative sum over iterated bubble diagrams, as depicted in Fig.~\ref{fig:prod8}(right). The channels included in this sum are summarised in Table~\ref{tab:channels}. For $1/2^-$ and $3/2^-$, we include all S-wave combinations of $\Lc\D\*$, $\S\*\D\*$, $N\jp$ and $N\eta_c$. (We include $N\eta_c$ as another possible final state of interest, though note that it features only in $1/2^-$, as we are assuming S-wave channels.) For $1/2^+$ the channels $\Lc(2595)\D$, $\Lc(2595)\D^*$ and $\Lc(2625)\D^*$ are included in S-wave, but the final state $N\jp$ is in P-wave (both $^2P_{1/2}$ and $^4P_{1/2}$), and we also include the related possible final state $N\eta_c(^2P_{1/2})$.

\begin{table}
    \centering
\begin{tabularx}{\textwidth}{Xl}
\hline
$1/2^-:$ &$\Lc\D,~\Lc\D^*,~\S\D,~\S\D^*,~\S^*\D^*,~N\jp,~N\eta_c$\\
$3/2^-:$ &$\Lc\D^*,~\S^*\D,~\S\D^*,~\S^*\D^*,~N\jp$\\
$1/2^+:$ &$\Lc(2595)\D,~\Lc(2595)\D^*,~\Lc(2625)\D^*,~N\jp(^2P_{1/2}),~N\jp(^4P_{1/2}),~N\eta_c(^2P_{1/2})$\\
\hline
\end{tabularx}
\caption{Channels included in the iterated sum over bubble diagrams.}
    \label{tab:channels}
\end{table}

We define the amplitudes $\mathcal{A}(^{2S+1}L_J)$ corresponding to a particular $\jpp$ spectroscopic state as follows:
\begin{align}
    \mathcal{A}(^2S_{1/2})&=b_1+g_1A_1+g_2A_2 \nonumber \\
    \mathcal{A}(^4S_{3/2})&=b_2+g_3A_3 \nonumber\\
    \mathcal{A}(^2P_{1/2})&=b_3+g_4A_4+g_5A_5+g_6A_6\nonumber \\
    \mathcal{A}(^4P_{3/2})&=b_4+g_4A_4^\prime+g_5A_5^\prime+g_6A_6^\prime.
    \label{eq:As}
\end{align}
The constituent amplitudes $A_i$ are computed from the triangle diagram with non-perturbative final state interactions, as described in the next sections, whereas the production couplings $g_i$ and background terms $b_i$ are treated as fit parameters. Notice in these expressions that amplitudes $A_i$ corresponding to the same final state are added coherently. Notice also that the $^2P_{1/2}$ and $^4P_{3/2}$ amplitudes have the same production couplings: they have the same production channels, but different spectroscopic final states.

The production couplings $g_{i}$, which account for the total strength of a given sub-amplitude, will be taken to be real because we assume that the dominant analytic structure for the amplitude is contained in $A_i$ (through the triangle diagram and final state interactions). 
The background terms $b_i$ account for the production of the given final state through mechanisms other than the triangle and final state interactions. Accordingly, these are complex valued, and for simplicity, we model these as constants. (We experimented with more sophisticated background models, for example with $s$-dependence, or additional incoherent contributions. Neither of these assisted in fit quality greatly. In our view overly strong $s$-dependence in background terms is to be avoided, and as we will see, is not necessary.)

Finally, the rate given by 
\begin{equation}
R = \int_{s_{12}(\textrm{min})}^{s_{12}(\textrm{max})} ds_{12}\,  \sum_{SLJ}|\mathcal{A}(^{2S+1}L_J)|^2
\end{equation}
is fit to the cos-weighted LHCb data. 

Our set-up is somewhat simpler than that of Du~\textit{et al.}~\cite{Du:2021fmf,Du:2019pij}. In particular, our constant complex background is to be compared with the more elaborate background of Du~\textit{et al.}, which contains a Breit-Wigner term. In our simplest cases (with only $1/2^-$ and $3/2^-$ channels), we fit four background terms (Re $b_1$, Im $b_1$, Re $b_2$, Im $b_2$), whereas the background in refs~\cite{Du:2021fmf,Du:2019pij} is parameterised by six constants.  

Another notable difference is that our model is more constrained on the production side. For example, in our simplest cases (with only $1/2^-$ and $3/2^-$ channels), we fit three production couplings ($g_1$, $g_2$, $g_3$), compared to seven in refs~\cite{Du:2021fmf,Du:2019pij}. The difference has a significant impact on the phenomenology. We remarked previously that an awkward consequence of Scenarios~A and B is the existence of $1/2^-$ and $3/2^-$ $\S^*\D^*$ states which are apparently not visible in the data. In the model of Du~\textit{et al.}, there are as many production couplings (seven) as there are $\S\*\D\*$ channels, resulting in enough parametric freedom to make the unwanted $\S^*\D^*$ states disappear from the $\jpp$ spectrum. In our model, with only three production couplings,  there is much less parametric freedom, and indeed we argue later that we would not be able to explain away the missing states in this way. Instead we avoid the problem by adopting parameter choices suitable for Scenario C, in which case the unwanted $1/2^-$ and $3/2^-$ $\S^*\D^*$ states simply do not bind, and so are absent for that reason.

\subsection{Amplitudes}
\label{sec:amplitudes}

\begin{figure}
    \centering
    \includegraphics[width=0.5\textwidth]{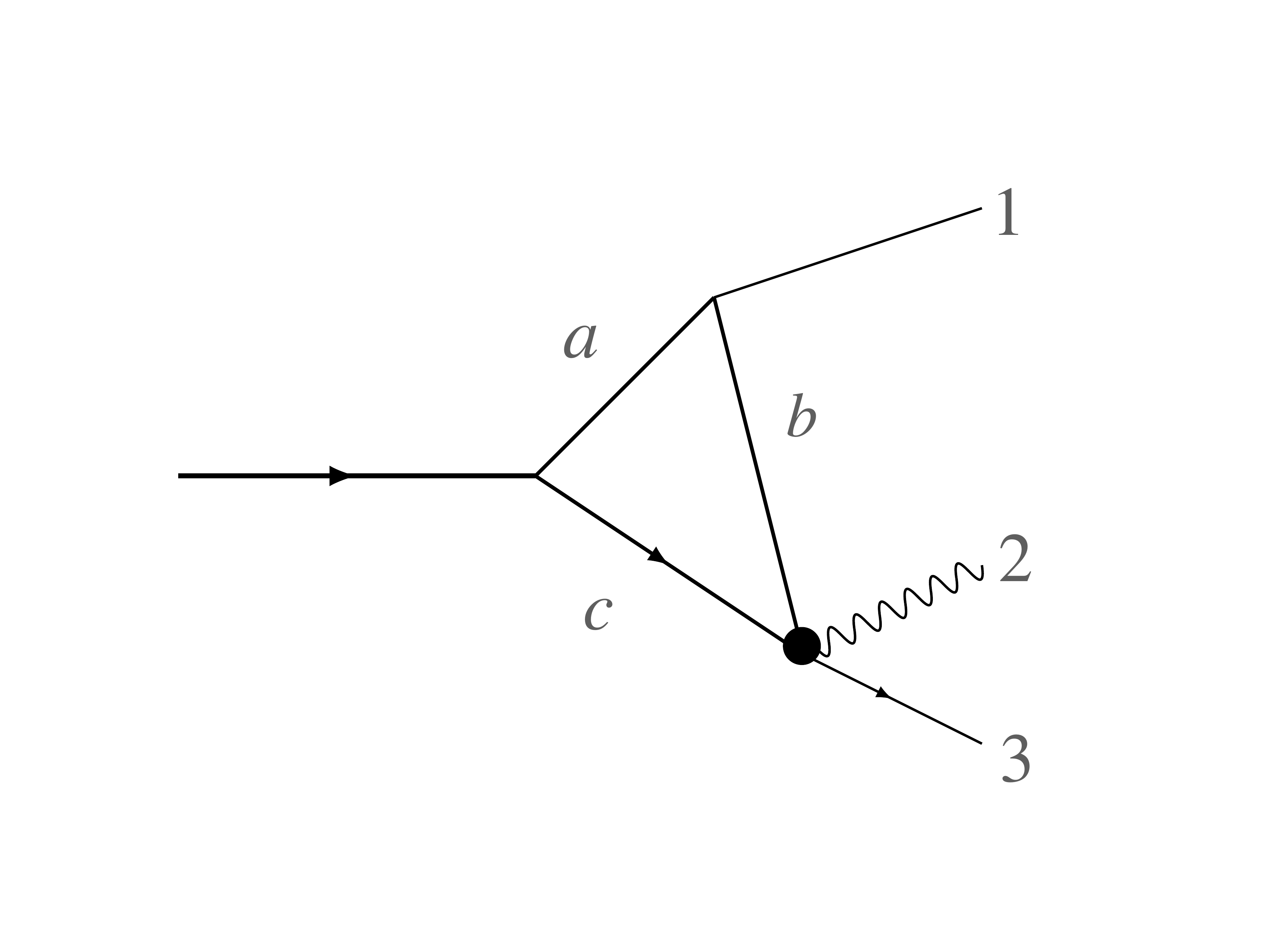}
    \caption{Generic diagram whose amplitude is discussed in the text.}
    \label{fig:prod3}
\end{figure}

Here we describe the calculation of the amplitudes $A_i$, with reference to the generic diagram in Fig.~\ref{fig:prod3}. We will use final states $1 = K^-$, $2= J/\psi$, $3= p$ and virtual states $a = ``D_s"$, $b = ``\bar D"$, and $c = ``\Lambda_c"$, where the quotation marks remind us that these refer to flavour only. As discussed above, we focus on low lying states with strong overlap with the initial and final states. Thus we consider $b = \bar{D}, \bar{D}^*$ and $c = \Lambda_c, \Lambda_c(2595), \Lambda_c(2625)$.

In principle the sum over $a$ can be done with a spectral integral. Again, we regard this as impractical; rather we assume that the sum is dominated by one resonance, typically the $D_s^*$ with a mass set to 2.112~GeV. We have found that changing the nominal $D_s$ mass has little effect on the rate once fit parameters are adjusted, \textit{unless} the mass is tuned to satisfy the Landau conditions that give rise to the triangle singularity~\cite{Landau}. With plausible $D_s$ masses this is only relevant for triangle diagrams with excited states in the triangle, specifically $\Lambda_c(2595)\bar{D}$ and related channels. Hence for the $1/2^+$ channel (only) we consider the case where a larger effective $D_s$ mass is used. More detail is provided in Section \ref{sect:case4}.

The amplitude is
 \begin{equation}
A =   \triangle_{abc}(s_{23}) \cdot  t_{bc:23}(s_{23}) \cdot Y^*_{L_f M_f}(\widehat{k_{23}}) F_{L_f}(k_{23}).
\label{eq:A}
\end{equation}
where $\triangle_{abc}(s_{23})$ is the amplitude for the triangle diagram, depending on the masses of $a,b,c$ and the invariant masses of the external legs (specifically $s_{23}$), and $t_{bc:23}(s_{23})$ is a (reduced) T-matrix which couples the state $bc$ in the triangle to the final state $23$, defined via
\begin{equation}
\langle \vec{p} L M; \alpha| T | \vec{p}' L' M'; \alpha'\rangle  \equiv Y_{LM}(\hat p) F_L(p) \cdot t_{\alpha \alpha'} \cdot Y^*_{L'M'}(\hat p') F_{L'}(p'),
\label{eq:t}
\end{equation}
where we are using $\alpha$ to label the scattering channels. 
Note that the separable form of the T-matrix permits the factorization shown in in Eq.\ref{eq:A}, wherein the initial form factor, $F_L$, is accounted for in the triangle integral.
 The  hadronic form factor $F_L$ is discussed further below. Finally, the T-matrix can be obtained by solving the Bethe-Heitler equation, $T=V+VGT$, using standard techniques.

We choose to write the triangle diagram with form factors that model the hadronic interactions at each of the vertices:
\begin{eqnarray}
\triangle_{abc}(s_{23}) &=& \int \dq F_\textrm{ew}(\bm{q})\, F_\textrm{3P0}(\bm{k} - x_1 \bm{q}) \, F_{L}(\bm{q} - x_2 \bm{k}) P_{L}(\bm{q} - x_2 \bm{k} ) \cdot \nonumber \\
&& [m_{\Lambda_b} - m_a - m_b - q^2/(2\mu_{ab}) + i \Gamma_{a}/2 + i\Gamma_b/2]^{-1} \cdot \nonumber \\
&& [m_{\Lambda_b}-E_1 - m_b - m_c - q^2/(2 m_c) - (\bm{q}-\bm{k})^2/(2 m_b) + i \Gamma_{b}/2 + i \Gamma_c/2]^{-1}.
\label{eq:tri}
\end{eqnarray}
A nonrelativistic form has been used for the diagram since all particles move at reasonably low speeds and because the loop integral is regulated beyond $\beta \sim \beta_{3P0} \sim \Lambda_{QCD}$. 
We have allowed for an angular momentum $L$ (normally 0 or 1) in the final state interaction vertex.
Also $x_1 = m_1/(m_1+m_b)$ and $x_2 = m_c/(m_b+m_c)$. The widths $\Gamma$ of the mesons $a$, $b$, and $c$ can have a strong attenuation effect on the amplitude. We include these as a constant imaginary part in the energy denominators, having found that this gives almost identical results to the use of dynamical widths. Finally, $\bm{k}$ is the momentum of the outgoing $(1)=K^-$ meson while its energy is $E_1$.

The electroweak vertex $F_\textrm{ew}$ is described by the matrix element $\langle ac|H_{EW}| \Lambda_b\rangle$ with $(ac) = D_s^* \Lambda_c$, $D_s \Lambda_c$, etc. We judge that modelling each of these cases individually is unjustifiable in view of the approximations being made and simply consider $(ac)= D_s^* \Lambda_c$. The leading relevant form factor has been estimated in the heavy quark limit by Guo and Kroll and can be approximated very well by the expression\cite{GK}

\begin{equation}
\xi(w) = \left(\frac{2}{1+w}\right)^8
\end{equation}
where $w = v_{\Lambda_b} \cdot v_{\Lambda_c}$ is the Isgur-Wise variable for $\Lambda_b \to \Lambda_c$. Including kinematic factors then gives the model vertex

\begin{equation}
F_\textrm{ew}(k) = \xi(w) \cdot k f_{D_s^*}
\end{equation}
where the last factor is the $D_s^*$ decay constant and $w^2 = 1 + k^2/m_{\Lambda_c^2}$.

The form factor $F_{3P0}$ is associated with the $(ab1)$ vertex, which corresponds to a strong decay such as $D_s^* \to K^-\bar{D}^{(*)0}$. We model this using the well-known ``3P0" model, which postulates that the interaction proceeds via quark-antiquark pair production with ${}^{(2S+1)}L_J = {}^3P_0$ quantum numbers \cite{Micu:1968mk,LeYaouanc:1972ae,Ackleh:1996yt,Barnes:1996ff,Barnes:2002mu,Burns:2007hk,Burns:2014zfa}. The vertex is obtained by integrating the relevant hadronic wavefunctions over the production operator. It is common to use simple harmonic oscillator wavefunctions with a scale that is fixed to reproduce the leading moments of the wavefunction as determined by a  quark model of choice. Generally this scale, called $\beta_{3P0}$ here, is the range 300-700 MeV, depending on the hadron~\cite{Barnes:2005pb}.

The strength of the vertex is absorbed into the production couplings $g_i$ which, as described previously, are fit parameters. We ignore possible polynomial dependence due to wavefunction nodes (there are none for the hadrons we consider), and we replace exponential dependence with the equivalent leading power law. Our final form is therefore

\begin{equation}
F_{3P0}(k) = \frac{x^\ell}{1+ x^2/12},\ \ x = k/\beta_{3P0}
\end{equation}
where $\ell$ is the angular momentum of the $(1b)$ system. We set $\beta_{3P0} = 500$ MeV in the following. This value is typical of quark model descriptions, as just mentioned, and we have found that our fits are not sensitive to variations in this parameter across a reasonable range.

We compute the integral in equation \eqref{eq:tri} numerically. In the special case that all form factors are set to unity and the particle widths are set to zero, the integral can be done analytically, and result is obtained in ref.~\cite{Guo:2019twa}. As a check on our code we have verified that our numerical result matches the analytical result in this limit.

The remaining vertex describes the final state interaction, $(bc) \to (23)$, for example $\Lambda_c \bar{D} \to J/\psi\, p$, manifest as a non-perturbative sum over bubble diagrams, as depicted in Fig.~\ref{fig:prod8}(right). Such interactions have been treated with effective Lagrangian or one-meson-exchange models \cite{Wu:2010vk,Roca:2015dva,He:2015cea,Chen:2015loa,Karliner:2015ina,Shimizu:2016rrd,Yamaguchi:2016ote,Yamaguchi:2017zmn,Shimizu:2017xrg,Shimizu:2018ran,Liu:2019zvb,PavonValderrama:2019nbk,Du:2019pij,Gutsche:2019mkg,Sakai:2019qph,Guo:2019kdc,He:2019ify,Liu:2019tjn,Chen:2019asm,Burns:2019iih,He:2019rva,Xiao:2019aya,Peng:2020xrf,Xu:2020gjl,Yalikun:2021bfm,Du:2021fmf}. Here we assume a separable form
\begin{equation}
\langle p L M; \alpha | V | p' L' M'; \alpha'\rangle = \lambda_{\alpha\alpha'} Y_{LM}(\hat p) F_L(p) \cdot Y^*_{L'M'} (\hat p') F_{L'}(p'),
\label{eq:V}
\end{equation}
corresponding to contact interactions with relative strengths $\lambda_{\alpha\alpha'}$  that are fixed by heavy quark symmetry (described below). 
This choice is made because the detailed form of the final state interactions is not known, and because it captures the relative strengths of the interactions in a simple way. Indeed, we expect that the important features of the final state interactions are the existence or absence of bound states, not the exact properties of these states. Lastly, a separable final state interaction Ansatz permits computing the full final state interactions in a simple fashion and permits easily folding them into the triangle diagram.
Thus, in view of the approximations made (and our goals) we judge it unnecessary to build a more elaborate final state interaction model. 

The hadronic form factor is modelled as
\begin{equation}
F_L(x) = \frac{x^{L}}{1+x^2}, \ \  x = p/\beta,
\end{equation}
where $\beta$ is a universal scale, $p$ is the relevant channel relative momentum,  and the numerator implements the expected angular momentum barrier. We apply this form factor at the ``$bc$'' vertex in the triangle diagram (Fig~\ref{fig:prod3}), and all vertices in the bubble diagrams (Fig~\ref{fig:prod8}), however for simplicity we do not apply the form factor in the $\jpp$ final state, as the momentum $p$ varies across the Dalitz plot.

The parameter was set to $\beta = 800$~MeV for the results reported here. This is a typical hadronic scale that leads to stable results of the correct magnitude. We have tested varying $\beta$ and find that its effects are largely absorbed in the final state interaction strengths, as expected.

The final state interactions are constrained to satisfy heavy-quark symmetry, and the corresponding channel coefficients $\lambda_{\alpha\alpha'}$, which we have obtained using standard means, are shown in  Tables \ref{tab:1}, \ref{tab:2}, and \ref{tab:3} for the $J^P= 1/2^-$, $3/2^-$, and $1/2^+$ systems, respectively. The parameters $A$, $B$, $C_{a,b}$, $D$, $E$, $F_{a,b}$, $G_{a,b}$ are contact terms which are fit to data, though not all of these are well-constrained by current experimental data (for example we will set $A=0$), and in some of the simpler scenarios we only need a subset of these parameters to achieve good fits.

The contact terms relevant to $1/2^-$ and $3/2^-$ channels have been discussed elsewhere in the literature -- see for example refs~\cite{Garcia-Recio:2013gaa,Xiao:2013yca,Xiao:2019aya,Sakai:2019qph,Du:2019pij,Du:2021fmf}. Our results for the $1/2^+$ matrix are new, and one should notice the similarity between the terms involving $F_{a,b}$ in the $1/2^+$ matrix and the corresponding terms involving $C_{a,b}$ in the $1/2^-$ matrix. The correspondence follows from replacing $\S$ with $\Lc(2595)$ (both $J=1/2$ states), replacing $\S^*$  with $\Lc(2625)$ (both $J=3/2$), and generalising the conservation of light quark spin to the conservation of light quark angular momentum.

Note that we are not explicitly including one-pion exchange in our final state interactions, and in this respect the model of Du~\etal\cite{Du:2019pij,Du:2021fmf} is more rigorous than ours. However we do not expect that the inclusion of one-pion exchange potentials will significantly alter our fits, since the pion-exchange contributions can to a certain extent be absorbed into an adjustment of the contact terms which, after all, are fit to data. This is particularly true of the central contributions to the potential, which follow the same pattern as the $B$ and $C_b$ terms in the $1/2^-$ and $3/2^-$ potential matrices. (More precisely, the pion-exchange potentials may be obtained from our Tables \ref{tab:1} and \ref{tab:2} by setting $A=C_a=0$, $C_b=-2B$, and replacing our separable potential with the Fourier transform of the scattering amplitude.) In this sense,  our model may be regarded as subsuming pion exchange (with, of course, a different spatial structure in the interactions). It is simply because the pion-exchange potentials respect heavy-quark symmetry.

Finally, we remark that the formulae above for the amplitude have been developed in the nonrelativistic formalism, hence the units of $\lambda_{\alpha \alpha'}$ are $-2$, three point functions are $-1/2$, and $-2$ for the amplitude for $\Lambda_b \to \jpp$. These amplitudes can be converted to relativistic conventions by multiplying by the usual factors of $\sqrt{2E}$. In our case, these are approximately $\sqrt{2M}$ and can be absorbed into the coupling constants. We have confirmed that  the appropriate factors of $\sqrt{E/M}$ make very little difference in our fits, with the chief effect being a slight enhancement at high energy that is easily countered by fit parameters.

\begin{table}
\begin{tabular}{l|cc|ccc|cc}
\hline\hline
$1/2^-$ & $\Lambda_c D$ & $\Lambda_c \D^*$  & $\Sigma_c \D$ & $\Sigma_c \D^*$ & $\Sigma_c^* \D^*$ & $ N J/\psi$ &$N \eta_c$\\
\hline
$ \Lambda_c \D$ & A & 0 & 0 & $\sqrt{3}B$ & $\sqrt{6}B$ & $\frac{\sqrt{3}}{2} D$ &$\frac{1}{2}D$\\
$ \Lambda_c \D^*$ & & A & $\sqrt{3}B$ & $-2B$ & $\sqrt{2}B$ & $-\frac{D}{2}$ &$\frac{\sqrt{3}}{2} D$\\
\hline
$\Sigma_c \D$ & &  & $C_a$ & $\frac{2}{\sqrt{3}} C_b$ & $-\sqrt{\frac{2}{3}} C_b$ & $-\frac{1}{2\sqrt{3}} E$  & $\frac{1}{2} E$ \\
$\Sigma_c \D^*$ & &  & & $C_a- \frac{4}{3}C_b$ & $-\frac{\sqrt{2}}{3} C_b$  & $\frac{5}{6} E$  & $-\frac{1}{2\sqrt{3}} E$ \\
$\Sigma_c^* \D^*$ & &  & & & $C_a- \frac{5}{3}C_b$ & $\frac{\sqrt{2}}{3} E$  & $\sqrt{\frac{2}{3}} E$ \\
\hline
$N J/\psi$ & & & & & & 0&0 \\
$N \eta_c$ & & & & & & &0 \\
\hline\hline
\end{tabular}

\caption{Contact terms in the $1/2^-$ channel.}
\label{tab:1}
\end{table}

\begin{table}
\begin{tabular}{l|c|ccc|c}
\hline\hline
$3/2^-$ & $\Lambda_c \D^*$ & $\Sigma_c^* \D$ & $\Sigma_c \D^*$ & $\Sigma_c^* \D^*$ & $N J/\psi$ \\
\hline
$ \Lambda_c \D^*$ & $A$ & $-\sqrt{3} B$ & $B$ & $\sqrt{5}B$ & $D$ \\
\hline
$\Sigma_c^* \D$ & & $C_a$ & $\frac{C_b}{\sqrt{3}}$ & $\sqrt{\frac{5}{3}} C_b$ & $-\frac{E}{\sqrt{3}}$ \\
$\Sigma_c \D^*$ & & & $C_a+\frac{2}{3}C_b$ & $-\frac{\sqrt{5}}{3} C_b$ & $\frac{E}{3}$ \\
$\Sigma_c^* \D^*$ & & & & $C_a-\frac{2}{3}C_b$ & $\frac{\sqrt{5}}{3}E$ \\
\hline
$N J/\psi$ & & & & & 0 \\ 
\hline\hline
\end{tabular}

\caption{Contact terms in the $3/2^-$ channel.}\label{tab:2}\end{table}

\begin{table}
\begin{tabular}{l|ccc|ccc}
\hline\hline
$1/2^+$ & $\Lambda_c(2595) \D$ & $ \Lambda_c(2595) \D^*$ & $\Lc(2625)\D^*$&$N\jp(^2P_{1/2})$&$N\jp(^4P_{1/2})$ &$N\eta_c(^2P_{1/2})$\\
\hline
$\Lambda_c(2595) \D$ & $F_a$ & $\frac{2}{\sqrt{3}} F_b$ &$-\sqrt{\frac{2}{3}}F_b$&$\frac{1}{6\sqrt 3}G_a-\frac{4}{3\sqrt 3}G_b$ & $\frac{1}{3}{\sqrt\frac{2}{3}}\left(G_a+G_b\right)$ & $\frac{1}{2}G_a$ \\
$\Lambda_c(2595) \D^*$ &  & $F_a- \frac{4}{3}F_b$ &$-\frac{\sqrt 2}{3}F_b$&$-\frac{5}{18}G_a-\frac{4}{9}G_b$ & $-\frac{10\sqrt 2}{18}G_a+\frac{\sqrt 2}{9}G_b$ & $-\frac{1}{2\sqrt 3}G_a$ \\
$\Lc(2625)\D^*$&&&$F_a-\frac{5}{3}F_b$&$-\frac{\sqrt 2}{9}G_a+\frac{2\sqrt 2}{9}G_b$&$-\frac{4}{9}G_a-\frac{1}{9}G_b$&$\sqrt{\frac{2}{3}}G_a$\\
\hline
$N\jp(^2P_{1/2})$ & && &  0& 0& 0\\
$N\jp(^4P_{1/2})$ & && & & 0& 0 \\ 
$N\eta_c(^2P_{1/2})$& && &  & & 0\\
\hline\hline
\end{tabular}

\caption{Contact terms in the $1/2^+$ channel.}\label{tab:3}
\end{table}

Before proceeding to the fits, it is interesting to consider higher order contributions to our production mechanism---effectively iterated triangles. These are discussed in the Appendix, where we conclude that their contributions are small with respect to the leading order considered here.

\section{Results}\label{sec:results}

\subsection{Fit Strategy}

We will fit our model to the $\jpp$ spectrum in $\Lambda_b\to J/\psi\, p \,K^-$ ~\cite{Aaij:2019vzc}, specifically the data set which has been weighted according to the cosine of the $P_c$ decay angle. We choose this particular data set because it enhances structure while retaining the full $Kp$ invariant mass phase space. The other LHCb data sets (raw data, or with a cut-off on the $Kp$ momentum) show similar features. Our fits cover the full range of possible invariant mass, in contrast to refs~\cite{Du:2021fmf,Du:2019pij}, which concentrate on the resonance region only, thereby essentially forcing a resonance interpretation of the features and neglecting important dynamics contained in the $\Lambda_c\*\D\*$ channels. 

We consider a series of five cases of increasing complexity, starting with minimal final state interactions and ending with interactions capable of generating poles for all the $P_c$ states including the $P_c(4457)$. The different cases, and their corresponding parameters, are summarised in Table~\ref{tab:fits}. Because fitting the entire data set is expensive, we chose to fit a representative subset of the data. The resulting values for chi-squared are reported in the the first row of the last section of the table. The next row contains chi-squared as computed over the full data set, while the final row reports chi-squared as computed over the resonance region, $M(J/\psi p) = 4.25$ -- $4.55$ GeV.

\begin{table}
    \centering
\begin{tabularx}{\textwidth}{Xlrrrrrr}
\hline
\input{fits}
\end{tabularx}
\caption{Model parameters for the various fits shown in Figs.~\ref{fig:fit1}-\ref{fig:fit5}.}

    \label{tab:fits}
\end{table}

Here we briefly introduce the cases and describe the fitting strategy.
In Cases 1 and 2 we only include the $1/2^-$ and $3/2^-$ channels. Case~1 is our minimal scenario, with no interactions among $\S\*\D\*$ channels ($C_a=C_b=0$); here we are considering the extent to which the data can be captured by kinematic features of the model, for example the direct coupling $\Lc\D\*\to\jpp$ of the triangle diagram to the final state (via the parameter $D$), and cusps due to  $\Lc\D\*\to\S\*\D\*\to\jpp$ (via $B$ and $E$). In Case~2 we switch on the couplings $C_{a,b}$ among the $\S\*\D\*$ channels, generating resonances to describe $P_c(4312)$, $P_c(4440)$, and the broad structure which was previously identified as $P_c(4380)$. As shown in Table~\ref{tab:fits}, we consider two variations (Cases 2a and 2b) with somewhat different parameters, but a similar physics interpretation and comparable fit quality. 

In the remaining cases we include not only  $1/2^-$ and $3/2^-$, but also $1/2^+$, and this has the largest impact on $P_c(4457)$, because of the proximity to $\Lc(2595)\D$ threshold. As shown in Fig.~\ref{fig:amps}, the inclusion of $1/2^+$ implies additional production diagrams due to $\Lc(2595)\D$, $\Lc(2595)\D^*$ and $\Lc(2625)\D^*$. In Cases 3 and 4 we consider the impact of the associated triangle diagrams, which are coupled to the  $\jpp$ final state via $G_{a,b}$. The difference between these cases is the ``$D_s$" mass used in the $1/2^+$ channel -- in Case~3 we use 2.112~GeV, whereas in Case~4 we adopt a heavier ``$D_s$" mass of 2.920~GeV, chosen to produce a logarithmic triangle singularity near the $\Lc(2595)\D$ threshold. Finally in Case~5 we revert to the standard ``$D_s$" mass of 2.112~GeV, but switch on final state interactions among $\Lc(2595)\D$, $\Lc(2595)\D^*$ and $\Lc(2625)\D^*$ (through $F_{a}$), leading to a resonance interpretation of $P_c(4457)$.

The parameters to fit are the contact terms ($B$, $C_a$, $C_b$, $D$, $E$, $F_a$, $G_a$, $G_b$), production couplings $g_i$, and backgrounds $b_i$. As shown in the table, the number of fit parameters in the cases we consider varies from 10 to 21. Note that we adopt $A=0$ in all cases, as our fits are not very sensitive to this parameter. (The contact term $A$ describes interactions among $\Lc\D\*$ channels, and the experimental data show no indication of strong attraction in such channels.)

In performing the fits we have found that standard minimisation algorithms cannot find a minimum, hence we have adopted a multi-step procedure wherein the contact terms were first adjusted manually, then the remaining parameters (production couplings and background terms) were fit to data. A difficulty in the more direct approach is that  changes in the parameters which couple the triangle diagrams or bubbles to the final states ($D$, $E$ and $G_{a,b}$) can be absorbed into changes in the associated production couplings $g_i$. Hence we prefer to fix these parameters according to other constraints, for example the known tiny branching fraction of $P_c$ states to $\jpp$ (which puts an upper limit on $E$). 

Also, the position of the $P_c(4312)$ and $P_c(4440)$ resonance peaks  is strongly influenced by $C_a$, $C_b$ and $B$, while the widths of the peaks -- particularly in the case of $P_c(4312)$ -- are constrained more strongly by $B$. (This is because the states, as discussed previously, decay dominantly to $\Lc\D^*$.) With these constraints in mind, it is convenient to fix these parameters first (by eye), and then separately fit the production and background terms.

We  remark that in choosing our ``best'' fits, we are (somewhat subjectively) aiming to get the best description of the data in the region of the $P_c$ peaks. In some cases, this is at the expense of the fit quality in regions of $\jpp$ invariant mass which are far from the $P_c$ peaks, and this to be expected, since our model has less leverage in those regions. We experimented with various fitting procedures, designed to tailor the fit to the region around the $P_c$ masses, but ultimately we find good results using a combination of manual tuning (for the contact terms) and algorithmic fitting (for everything else).

\subsection{Case~1}

The first case examines the degree to which the simple kinematical features of the model---the production triangle and threshold cusps---can explain the data. Thus we consider the $J^P = 1/2^-$ and $3/2^-$ channels only and set $C_a=C_b=0$, while retaining couplings between the initial and final state ($B$, $D$, $E$). The $D_s$ mass is set to 2.112 GeV. 

\begin{figure}[ht]
    \centering
    \includegraphics[width=\textwidth]{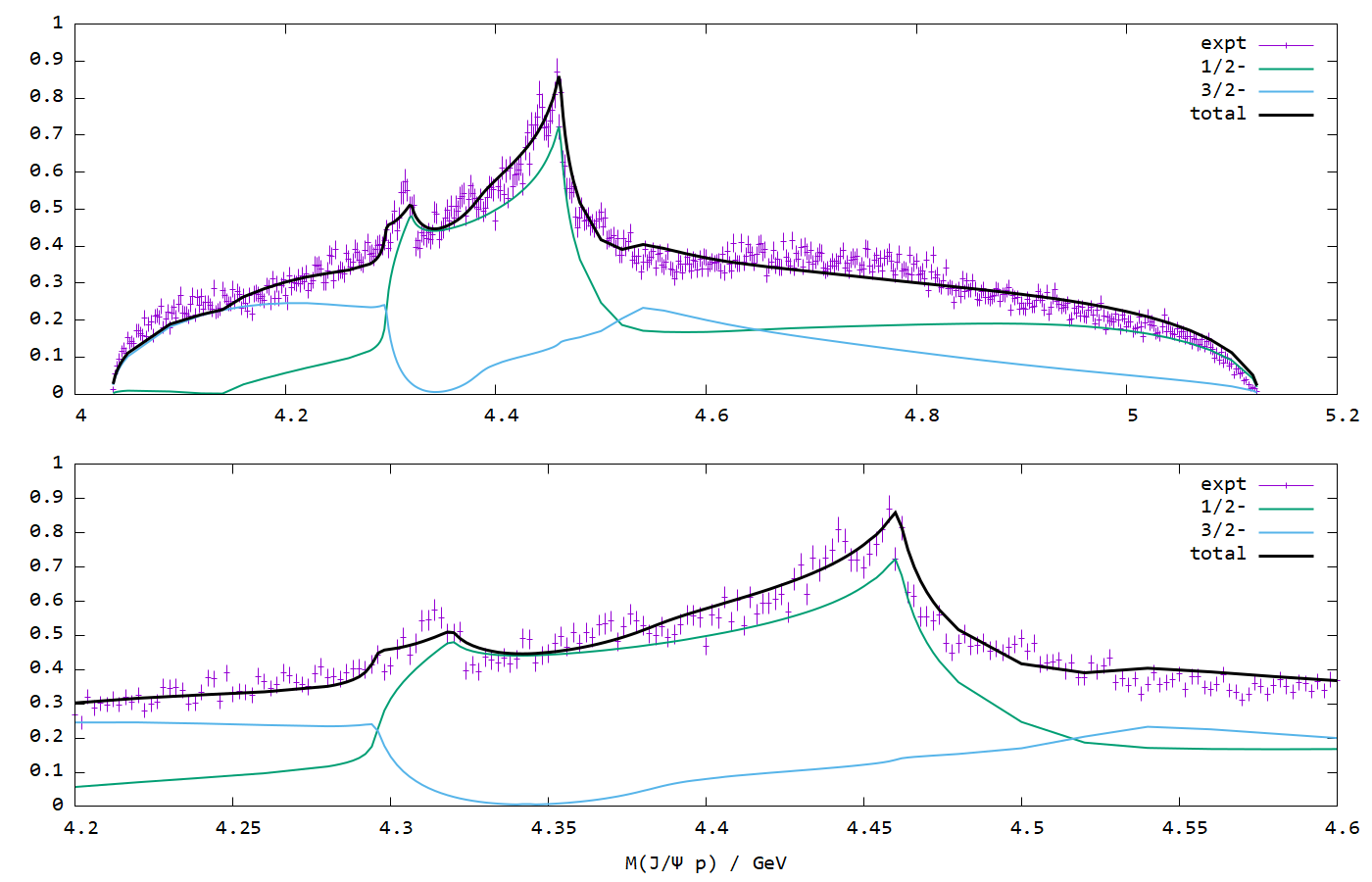}
    \caption{The $\jpp$ invariant mass spectrum in Case~1, in which $P_c(4457)$ is a $\S\D^*\to\jpp$ cusp.}
    \label{fig:fit1}
\end{figure}

With this set up the signal comes from the (off-singularity) triangle in  $\Lc\D\*\to \jpp$ (via $D$) and the triangle and threshold cusps in $\Lc\D\*\to\S\*\D\*\to \jpp$ (via $B$ and $E$). The result is displayed in Fig.~\ref{fig:fit1}. 

It is immediately clear that even this simple model captures the majority of the bulk properties of the data set. In more detail, a shoulder appears at $\Lambda_c\bar{D}$ threshold in $1/2^-$. This combines with the low-lying $3/2^-$ spectrum that follows phase space in this region. The next feature occurs near the $P_c(4312)$ peak where we see that the  $3/2^-$ dips down due to interference above  $\Lambda_c\bar{D}^*$ and $J^P =1/2^-$  increases above $\Sigma_c\bar{D}$ due to the threshold opening.  At higher energy, we see a threshold cusp  due to $\Sigma_c\bar{D}^*$ coupling to the production triangle via the coupling $B$.  This does an excellent job of describing the $P_c(4457)$ peak---including the unusually sharp drop on the high energy side of the peak. We consider this strong support for the hypothesis that the $P_c(4457)$ can be explained by ``kinematical" effects. Ref.~\cite{Kuang:2020bnk} reaches a similar conclusion.

That this simple model can reproduce such intricate features is noteworthy, and draws attention to a fundamental difference between our approach and that of Du~\etal\cite{Du:2021fmf,Du:2019pij}. In our case, the main features in the data are captured through a coherent combination of contributions from the triangle diagrams (involving $\Lc\D\*$ states), re-summed bubble diagrams (including $\S\*\D\*$ channels), and a simple constant background. By contrast, in refs ~\cite{Du:2021fmf,Du:2019pij}, the signal (due to $\S\*\D\*$ interactions) and background (which is more intricate than ours, but smooth) are combined incoherently; by construction this leads to a resonance interpretation of any features which deviate from the background, but as is apparent in Fig.~\ref{fig:fit1}, even dramatic peaks need not necessarily be associated with resonances.

Note that the values we adopt for $B$, $D$ and $E$ (Table~\ref{tab:fits}) are not well-constrained in this case, so for convenience we adopt values which are the same or similar to those used in subsequent (more tightly constrained) cases.

\subsection{Case~2}

Although encouraging, Case~1 makes it clear that additional dynamics is required to explain the sharpness of the $P_c(4312)$ and $P_c(4440)$ peaks, as well as the structure around 4380~MeV. We therefore follow the strategy developed above and turn on final state interactions in the $\Sigma_c^{(*)}\bar{D}^{(*)}$ sector by adjusting $C_a$ and $C_b$, adopting negative values for both in order to implement our preferred Scenario~C. We consider two implementations for this case, one with $C_a=-14.8$~GeV${}^{-2}$ and $C_b=-8.0$ GeV${}^{-2}$ and the other with $C_a=-14.0$~GeV${}^{-2}$ and $C_b= -9.8$~GeV${}^{-2}$. We have found that both options yield  resonance positions  suitable for making fits. 

Compared to the previous case, the choice of values for the other contact terms ($B$, $D$ and $E$) has more significance, and we discuss this further below. As mentioned, once we have fit the contact terms -- in particular, to generate peaks associated with $P_c(4312)$ and $P_c(4440)$ -- we then fit the remaining parameters (production couplings and background terms) to data. Note that in this model the $J^P=1/2^+$ channel is still set to null.

\begin{figure}[ht]
    \centering
    \includegraphics[width=\textwidth]{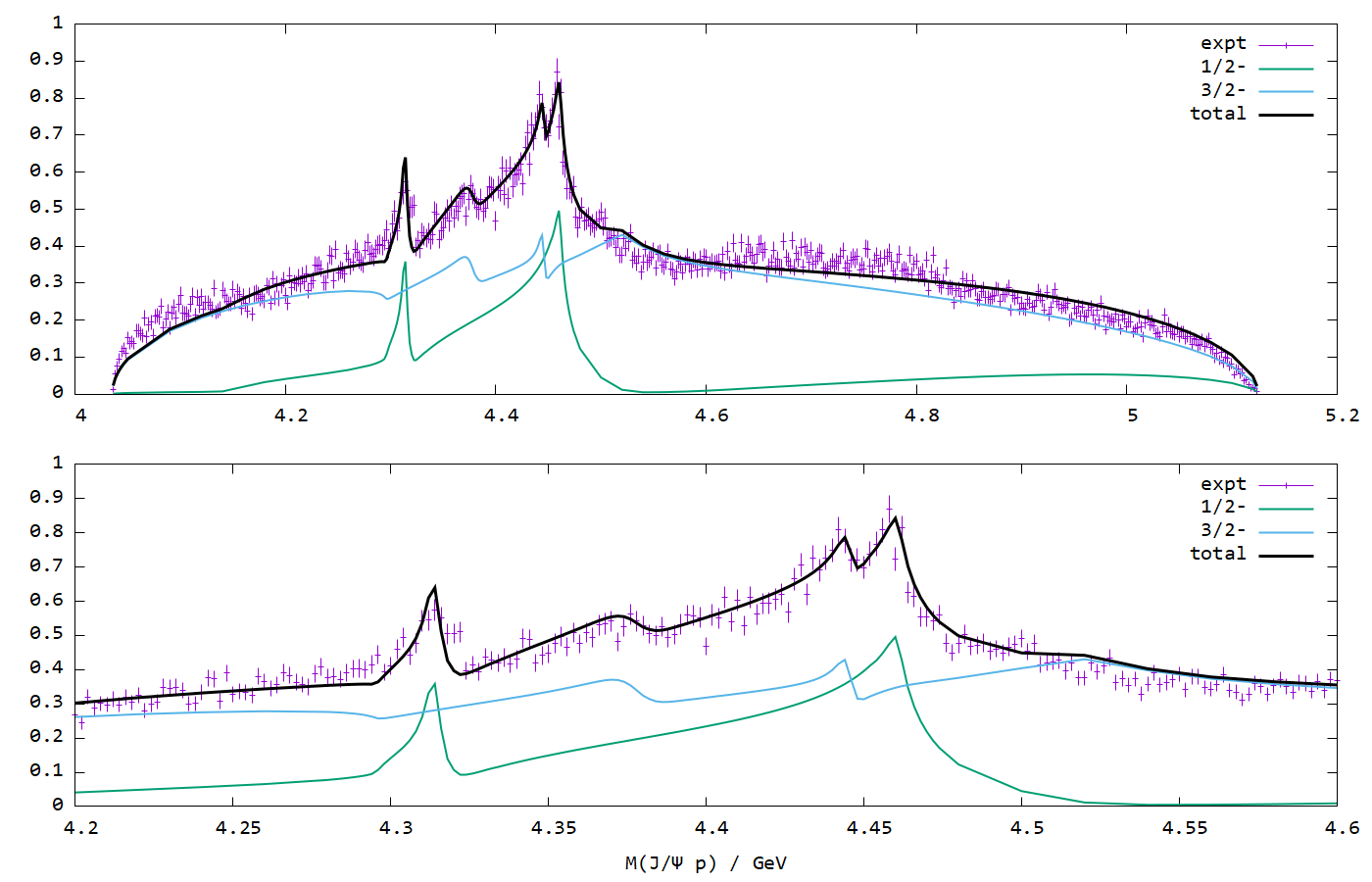}
    \caption{The $\jpp$ invariant mass spectrum in Case~2a, where in addition to the kinematical features included in Case~1, there are final state interactions in the $\S\*\D\*$ sector, leading to resonances for $P_c(4312)$, $P_c(4380)$ and $P_c(4440)$.}
    \label{fig:fit2a}
\end{figure}

\begin{figure}
    \centering
    \includegraphics[width=\textwidth]{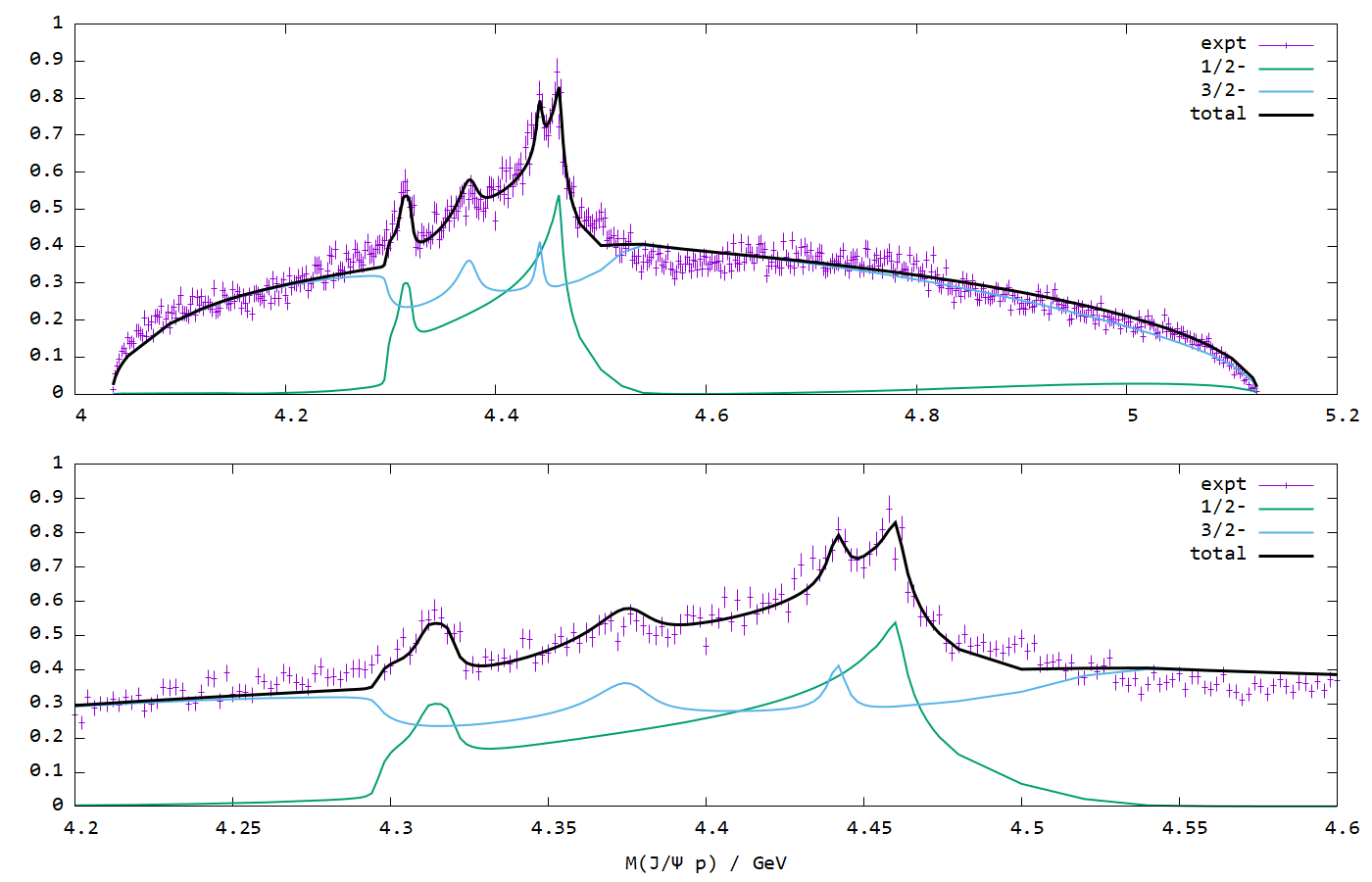}
    \caption{The $\jpp$ invariant mass spectrum in Case~2b, which is very similar to Case 2a but with somewhat different values for the contact terms (see Table~\ref{tab:fits}). Notice in particular the larger width for $P_c(4312)$, resulting from the larger value for $B$.}
    \label{fig:fit2b}
\end{figure}

Results for the two implementations (Cases 2a and 2b) are shown in Figs. \ref{fig:fit2a} and \ref{fig:fit2b}, respectively. Focussing attention on Fig. \ref{fig:fit2a} for the moment (Case~2a), one observes that the desired effects have been achieved, namely the $P_c(4312)$ peak is reproduced as a $1/2^-$ resonance, the $P_c(4380)$ appears as a $J^P=3/2^-$ resonance, as does the $P_c(4440)$. The $P_c(4457)$ remains a threshold cusp in $\Sigma_c\bar{D}^*$ as in the previous case. Notice that the peak at 4312~MeV in $1/2^-$ encourages the fit to shift the background from $J^P=1/2^-$ to $3/2^-$. The last feature to note is the broad rise near $\Sigma_c^*\bar{D}^*$ in $J^P=3/2^-$ (which is also present in Case~1). This is due to the strong diagonal interaction in this channel, although of course, it is weaker than $\Sigma_c^*\bar{D}$ and $\Sigma_c\bar{D}^*$ because $C_b$ is chosen to be negative.

As shown in Fig.~\ref{fig:fit2b}, Case~2b has similar features to Case~2a, as might be expected, with the chief difference being the shape of the $\Sigma_c^*\bar{D}^*$ rise in $J^P= 3/2^-$. Differences in the widths of the peaks are associated with the choice of $B$, as discussed below.

Notice that Cases 2a and 2b nicely capture the broad feature near 4380~MeV, a structure which was previously identified as the $P_c(4380)$~\cite{Aaij:2015tga,Aaij:2016phn}. As noted previously, a resonance around this mass is automatic in molecular models with heavy quark symmetry, once the potential is tuned to generate the $P_c(4312)$ peak.

Another notable feature is that the model nicely captures the relative widths of the $P_c(4380)$ peak (broad) compared to those of $P_c(4312)$ and $P_c(4440)$ (narrow). This follows naturally from the intrinsic widths of the dominant constituents of the states, which enter into the imaginary part of the energy denominators in equation~\eqref{eq:tri}. Since $\Sigma_c^*$ ($\Gamma =15$~MeV) is significantly broader than  $\Sigma_c$ ($\Gamma = 1.86$~MeV), the $P_c(4380)$ peak ($\S^*\D$) is naturally much broader than $P_c(4312)$ and $P_c(4440)$ ($\S\D$ and $\S\D^*$). Indeed we have verified, by varying the $\S^*$ width, that this is the origin of the effect. The simple explanation for the relative widths of the $P_c$ states lends support to the molecular scenario.

As mentioned above, the choice of values for the contact terms $B$, $D$ and $E$ has more significance than in the previous case. We argued in our previous paper \cite{Burns:2021jlu} that $P_c(4312)$ decays dominantly to $\Lc\D^*$, hence we expect (and have observed) that the width of the $P_c(4312)$ peak is strongly correlated with $B$, which controls the $\S\D\to\Lc\D^*$ coupling. (For a similar reason, the widths of $P_c(4380)$ and $P_c(4440)$ are also correlated with $B$, but less strongly, because of the smaller $\Lc\D\*$ branching fractions.) We have found that taking $B\approx 4\div 6$~GeV${}^{-2}$ generates a suitable $P_c(4312)$ width. The effect of $B$ on the widths is apparent when comparing Figs~\ref{fig:fit2a} and \ref{fig:fit2b}, which have  $B=4$~GeV${}^{-2}$ and $B=6$~GeV${}^{-2}$, respectively. 

In Section~\ref{sec:amplitudes}, we remarked that the algebraic structure of the one-pion exchange potential is reproduced with our potentials, taking $C_b=-2B$. It is noteworthy that the values of $B$ and $C_b$ preferred by our fit are roughly consistent with this relation (see Table~\ref{tab:fits}).

The parameter $B$ influences not only the shapes, but also the positions of the peaks, particularly because it controls the coupling between the nearby thresholds $\S\D$ and $\Lc\D^*$, which is partly responsible for the attraction generating $P_c(4312)$. This explains why -- comparing Cases~2a and 2b in Table~\ref{tab:fits} -- we find that an increase of $B$ must be compensated by a decrease in the magnitude of $C_a$.

Additionally, we note that the fit is not sensitive to $D$ and $E$ separately, but only the ratio $D/E$, for the following reason. The $\jpp$ final state can arise through $\Lc\D\*\to\jpp$ or $\S\*\D\*\to\jpp$, which scale with $D$ and $E$, respectively. It follows that the amplitude is a coherent sum of two terms, which scale with $D$ and $E$ respectively, provided that $D$ and $E$ are perturbatively small, meaning that  multiple rescatterings (such as $\Lc\D\*\to\jpp\to \Lc\D\*\to\jpp$) make negligible contribution. In this case the shape of the fit is sensitive only to the ratio $D/E$, and we have verified that this is true over a large range of $D$ and $E$, including the range of values which are allowed by phenomenology (discussed below). In practice it means that if we re-scale the contact terms $D$ and $E$ by a common factor, we get an essentially identical fit, but with the production couplings $g_i$ having absorbed the re-scaling. 

Hence in order to fix $D$ and $E$ (not just their ratio), we need additional input. It has been established experimentally that the branching fractions $P_c\to\jpp$ are very small, which implies an upper limit on $E$. In our previous paper \cite{Burns:2021jlu} we argued that $\mathcal B(P_c(4312)\to\jpp)$ must be less than a few parts in $10^{-3}$, but not significantly less -- this is in order to be consistent with experimental upper limits on the photoproduction cross sections \cite{GlueX:2019mkq,Joosten2021}, without implying an unrealistic production branching fraction. We may estimate the ratio of $B$ and $E$ from the potential matrix elements
\begin{align}
    \left|\frac B E \right|=\frac 1 6\sqrt{\frac{\mathcal{B}(P_c(4312)\to\Lc\D^*)}{\mathcal{B}(P_c(4312)\to\jpp)}},
    \label{eq:BE}
\end{align}
where we have ignored differences due to phase space. From the above arguments, we estimate $\mathcal B(P_c(4312)\to\jpp)\approx 10^{-3}$ and, from our previous analysis~\cite{Burns:2021jlu}, $\mathcal B(P_c(4312)\to\Lc\D^*)=59\div87\%$. Together these imply $|B/E|\approx 4\div 5$. Having fixed $B$ as outlined above, we adopt $E=1.0$~GeV$^{-2}$ as a value in the suitable range. We will later confirm, through an analysis of the T-matrix poles, that with this value we obtain the branching fraction  $\mathcal B(P_c(4312)\to\jpp)\approx 10^{-3}$, which is consistent with the above experimental constraints.

Having fixed $E$, a suitable value for $D$ is constrained by the sensitivity of the fit to the ratio $D/E$, as outlined above. As shown in Table~\ref{tab:fits}, we settle on values of $D=0.6$ or $0.7$~GeV$^{-2}$ as giving good results -- values significantly out of this range give features with the wrong shape. Since $D$ and $E$ respectively control the coupling of $\Lc\D\*$ and $\S\*\D\*$ to $\jpp$, it is noteworthy that the fit settles on values of comparable magnitude for these parameters. This is reassuring, considering the underlying similarity (in a quark model sense) of $\Lc$ and $\S\*$, and it supports out perspective, emphasised throughout the paper, that $\Lc\D\*$ channels are equally as important as $\S\*\D\*$ channels in describing the experimental data.

\subsection{Case~3}

In the remaining cases we introduce the $J^P=1/2^+$ channel, turning on the amplitudes $\mathcal{A}(^2P_{1/2})$ and $\mathcal{A}(^4P_{1/2})$ in eqn \eqref{eq:As}. The production in these amplitudes is via triangle diagrams with $\Lc(2595)\D$, $\Lc(2595)\D^*$ and $\Lc(2625)\D^*$ (Fig.~\ref{fig:amps}). 

In this first implementation of the $1/2^+$ channel, we keep the same ``$D_s$'' mass as used in the previous cases. The contact terms associated with the $1/2^-$ and $3/2^-$ channels ($B$, $C_a$, $C_b$, $D$, $E$) are set to the same values used in Case~2a, reproducing the peak positions for $P_c(4312)$, $P_c(4380)$ and $P_c(4440)$. As for the contact terms associated with the $1/2^+$ channel ($G_a,G_b$), we find that the fit is somewhat sensitive to the ratio $G_a/G_b$, but essentially insensitive under a common re-scaling of $G_a$ and $G_b$, as such re-scaling is absorbed into the production couplings $g_{4,5,6}$ which are fit to data. The situation is similar to the $1/2^-$ and $3/2^-$ channels (described previously), where $\jpp$ spectrum does not constrain $D$ and $E$ directly, but only their ratio $D/E$. A notable difference is that in that case, additional experimental input could be used to constrain the magnitude of $E$ (hence $D$). In the case of $G_a$ and $G_b$ there is no such constraint. We adopt the value $G_a=0.3$~GeV$^{-2}$ -- which has no particular significance -- and explore different values for $G_b$.

Regarding the background, we experimented with using the same treatment as in the previous cases, namely we fit $b_1$ and $b_2$, but keep $b_3=b_4=0$, so that we are assuming effectively that the background is dominated by $\jpp$ S-wave channels ($1/2^-$ and $3/2^-$). The resulting fit is, qualitatively, no better than those of previous cases, despite having more fit parameters. 

\begin{figure}
    \centering
    \includegraphics[width=\textwidth]{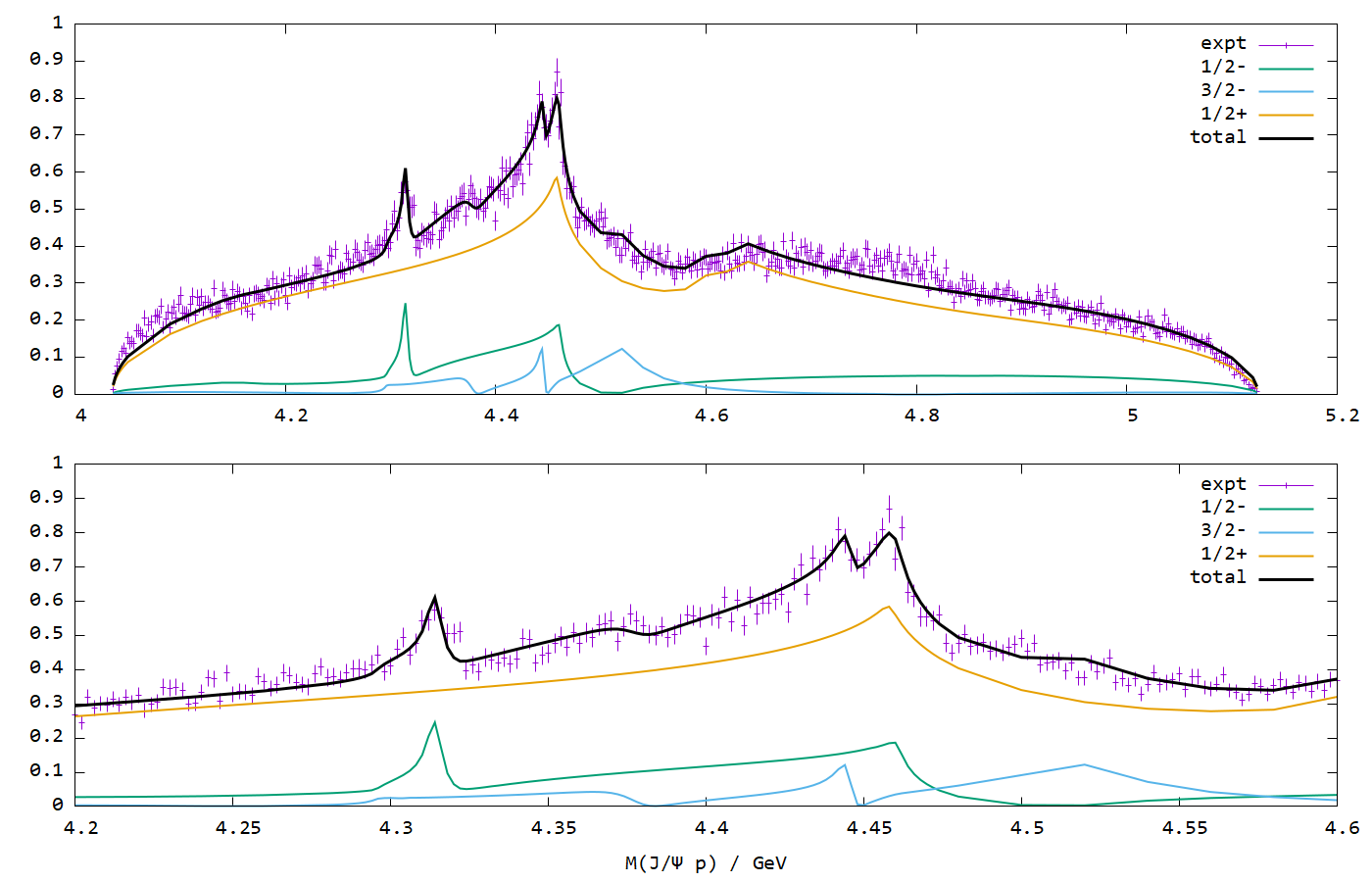}
    \caption{The $\jpp$ invariant mass spectrum in Case~3, in which the $1/2^+$ channel features prominently. The fit quality is better than previous cases, particularly in the higher mass region -- but the dominance of $\jpp$ in P-wave is unnatural.}
    \label{fig:fit3}
\end{figure}

Hence for illustration we performed another fit with non-zero background in all channels ($1/2^-$, $3/2^-$, $1/2^+$), namely we fit all of $b_1$, $b_2$, $b_3$ and $b_4$. This introduces two additional complex parameters, and unsurprisingly, results in an improved fit, shown in Fig.~\ref{fig:fit3}. In particular, we notice in comparison to previous cases that the fit is much improved in the region above the $P_c$ states, because of the leverage associated with the $\Lc(2595)\D$, $\Lc(2595)\D^*$ and $\Lc(2625)\D^*$ degrees of freedom (note the position of thresholds in Fig.~\ref{fig:thresholds}). The amplitudes $A_{4,5,6}^{(\prime)}$ have cusps at the $\Lc(2595)\D$, $\Lc(2595)\D^*$ and $\Lc(2625)\D^*$ thresholds, and their relative contributions are determined by the fit, as each has an associated production coupling ($g_{4,5,6}$). The $\Lc(2595)\D^*$ and $\Lc(2625)\D^*$ cusps are particularly useful in capturing the rise in data above 4.6~GeV.

For this particular fit we are using $G_b=0.2$~GeV$^{-2}$, but this is not well-constrained by data.

While the fit is very good, it is noteworthy that the $1/2^+$ amplitude has absorbed much of the background strength in the reaction. Although the data is agnostic to this possibility, we find it somewhat implausible physically, because it implies a substantial coupling to $\jpp$ in P-wave.

\subsection{Case~4}
\label{sect:case4}

We now consider the implications of varying the mass of the ``$D_s$" meson in the production diagrams, in order to investigate a possible role for the logarithmic triangle singularity. With reference to Fig.~\ref{fig:prod8}, the idea is that by tuning the $D_s$ mass, we may generate a logarithmic singularity in the $\jpp$ spectrum at (or above) the threshold for $\Lc\D\*$, $\Lc(2595)\D\*$ or $\Lc(2625)\D$. Of course we are mainly interested in the $\Lc(2595)\D$ case, as the strong and sharp $P_c(4457)$ peak coincides exactly with the threshold. This mechanism was suggested after the initial discovery of $P_c$ states \cite{Liu:2015fea}, and was confronted with experimental data in the more recent LHCb paper~\cite{Aaij:2019vzc}, with the conclusion that further analysis was warranted in future amplitude analyses. Other authors have considered different triangle diagrams for the $P_c$ states, involving other combinations of hadrons~\cite{Guo:2015umn,Mikhasenko2015,Guo:2016bkl,Bayar:2016ftu,Guo:2019twa}. Unlike the case we consider, these diagrams either violate isospin or are colour-suppressed.

Solving the  non-relativistic version of the Landau equations~\cite{Guo:2019twa}, we can determine the ``$D_s$" mass that is required to induce the logarithmic triangle singularity. For diagrams with $\Lc(2595)\D$ (amplitude $A_4^{(\prime)}$) we find $2.916\textrm{ GeV}<m_{D_s}<3.024\textrm{ GeV}$. Considering that there are several known $D_s$ mesons in a similar mass range \cite{ParticleDataGroup:2020ssz},  it is plausible that the logarithmic triangle singularity is indeed playing a role. Although we may expect the production of such a highly excited state to be suppressed in the weak decay, its contribution to the spectrum could nonetheless be significant because of the dramatic enhancement of the triangle amplitude because of the logarithmic singularity. We will adopt a ``$D_s$" mass of 2.92~GeV, namely at the lower end of the allowed range. Taking a much larger ``$D_s$" mass moves the triangle peak above $\Lc(2595)\D$ threshold, in conflict with data.

Interestingly, if we repeat the calculation for diagrams with $\Lc\D$ or $\Lc\D^*$ (amplitudes $A_{1,2,3}$) we find that the requisite $D_s$ mass is well above the masses of any known $D_s$ mesons. Hence we do not expect the triangle singularity to play a role in these cases, which is consistent with the absence of any prominent peaks in the data at the $\Lc\D$ and $\Lc\D^*$ thresholds.

Hence in Case~4 we use a ``$D_s$" mass of 2.920~GeV for the $1/2^+$ channel, and ``$D_s$" mass of 2.112~GeV for the $1/2^-$ and $3/2^-$ channels. (We experimented with summing over contributions from both masses for all channels, but this introduces many more free parameters, without an appreciable improvement in the fit.) For the background, we revert to the previous assumption (as in Cases 1 and 2) that it is dominated by $\jpp$ in S-wave (hence we fit $b_1$ and $b_2$, but keep $b_3=b_4=0$). For the contact terms, we fix $B$, $D$, $E$ and $G_a$ to the same values as the previous case, but vary the other terms, settling on values which are only slightly different to the previous case (but $G_b$ in particular is not tightly constrained). 

\begin{figure}
    \centering
    \includegraphics[width=\textwidth]{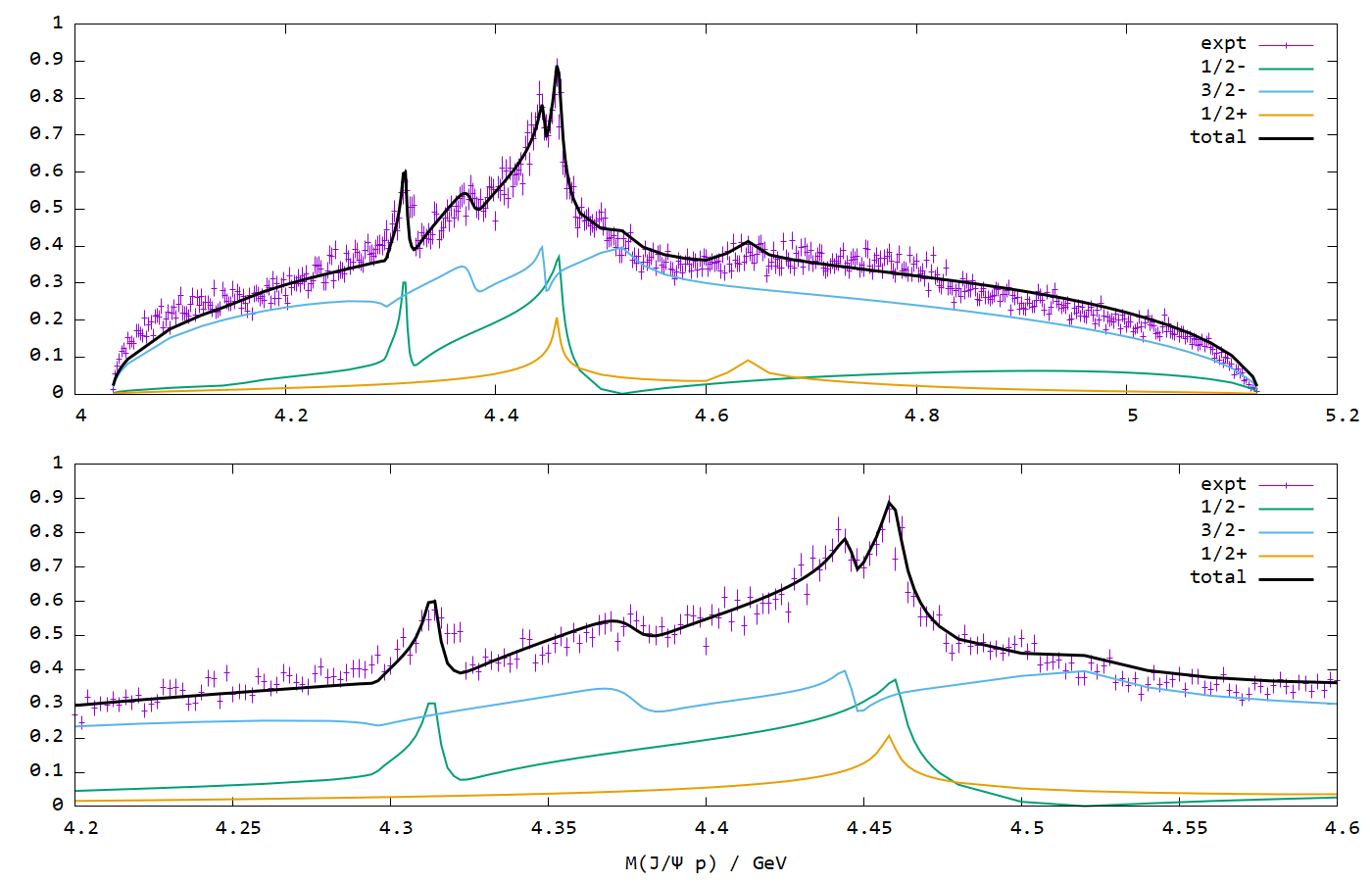}
    \caption{The $\jpp$ invariant mass spectrum in Case~4, in which the triangle singularity in $1/2^+$ gives a narrow peak at $\Lc(2595)\D$  threshold, making a significant contribution to the $P_c(4457)$ signal.}
    \label{fig:fit4}
\end{figure}

The result is shown in Fig.~\ref{fig:fit4}. The fit is of comparable quality to Case~3, but is physically more plausible, because the spectrum overall is dominated by $\jpp$ in S-wave ($1/2^-$ and $3/2^-$), rather than P-wave ($1/2^+$). The features in the $1/2^+$ channel are sharply localised around the corresponding thresholds, which is a consequence of their origin in the logarithmic singularity. The triangle diagram with $\Lc(2595)\D$ gives a sharp feature which contributes to $P_c(4457)$, while the $\Lc(2625)\D^*$ diagram nicely reproduces the rise in data above 4.6~GeV. 

It is notable that the couplings $g_{4,5,6}$ are significantly smaller in Case~4 compared to Case~3 (see Table~\ref{tab:fits}), which is to compensate the significant enhancement in the amplitudes $A_{4,5,6}^{(\prime)}$ resulting from the triangle singularity. This underlines our previous point that even if the production of a highly excited $D_s$ meson is suppressed at the weak vertex, its contribution to the spectrum can still be significant, due to the triangle enhancement.

In comparison to Case~2, the fit in Case~4 is of comparable quality in the region of the $P_c$ states, but better at higher mass, although at the cost of five more parameters.

\subsection{Case~5}

The last case examines the effect of constructing a $P_c(4457)$ resonance in the $1/2^+$ channel, which can be achieved by setting the coupling $F_a$ (see Table \ref{tab:3}) negative and sufficiently large. This scenario most closely reflects the suggestion of  Ref.~\cite{Burns:2019iih}, namely that an S-wave $\Lambda_c(2595)\bar D$ attraction provides a viable interpretation of the $P_c(4457)$. In that paper the attraction is attributed to one-pion exchange diagrams $\Lc(2595)\D\to\S\D^*$, although we noticed that with the potential strength fixed from experimental couplings, this diagram alone is insufficient to cause binding, a point examined in the subsequent, more detailed studies of Yalikun~\etal\cite{Yalikun:2021bfm}. Our approach in ref.~\cite{Burns:2019iih} was to enhanced the coupling strength until the desired binding was achieved, arguing that it could be justified by assuming some additional attraction due to unknown short-distance physics. In this paper, we ignore the meson exchange potential entirely, and include only the short-distance terms, following the same approach as we used in the $1/2^-$ and $3/2^-$ channels.

Unlike Case 4, we are no longer relying on the triangle singularity to account for $P_c(4457)$, so we revert to the ``$D_s$" mass of 2112 MeV as in cases 1, 2, and 3. For the background, we assume (as usual) that it is dominated by $\jpp$ in S-wave (fitting $b_1$ and $b_2$ only). We set most of the contact terms to the same values which we used in Case~2a, finding that there is no noticeable improvement in the results if we vary these.

When introducing resonances into the $1/2^-$ and $3/2^-$ channels, we found that the contact terms are quite well-constrained: for a given $B$, we have essentially two parameters ($C_a$ and $C_b$), which are constrained to fit the positions of three peaks. In the $1/2^+$ case, by contrast, we have an analogous two parameters ($F_a$ and $F_b$), but only one peak, hence much more parametric freedom. Moreover, the $P_c(4312)$, $P_c(4380)$ and $P_c(4440)$ peaks are far enough from any production thresholds that they can solely be attributed to resonances, which tightly constrains their parameterisation. By contrast, the $P_c(4457)$ peak has a contribution from the $\S\D^*$ and $\Lc(2595)\D$ triangle cusps, even before introducing attractive $\Lc(2595)\D$ interactions, so any resonant contribution is not well-constrained.

Given these limitations, for simplicity we fix the $1/2^+$ attraction  by setting $F_b=0$ and varying $F_a$. We achieve good fits with $F_a$ in the range of approximately $-6$ to $-16$~GeV$^{-2}$. As we increase the magnitude of $F_a$, the structure at $\Lc(2595)\D$ threshold evolves from a cusp (due to its production in the triangle) into a sharper feature which is characteristic of the onset of $\Lc(2595)\D$ binding. If we make $F_a$ too large, this sharp peak moves off below the $\Lc(2595)\D$ threshold, corresponding to increased binding energy.

\begin{figure}
    \centering
    \includegraphics[width=\textwidth]{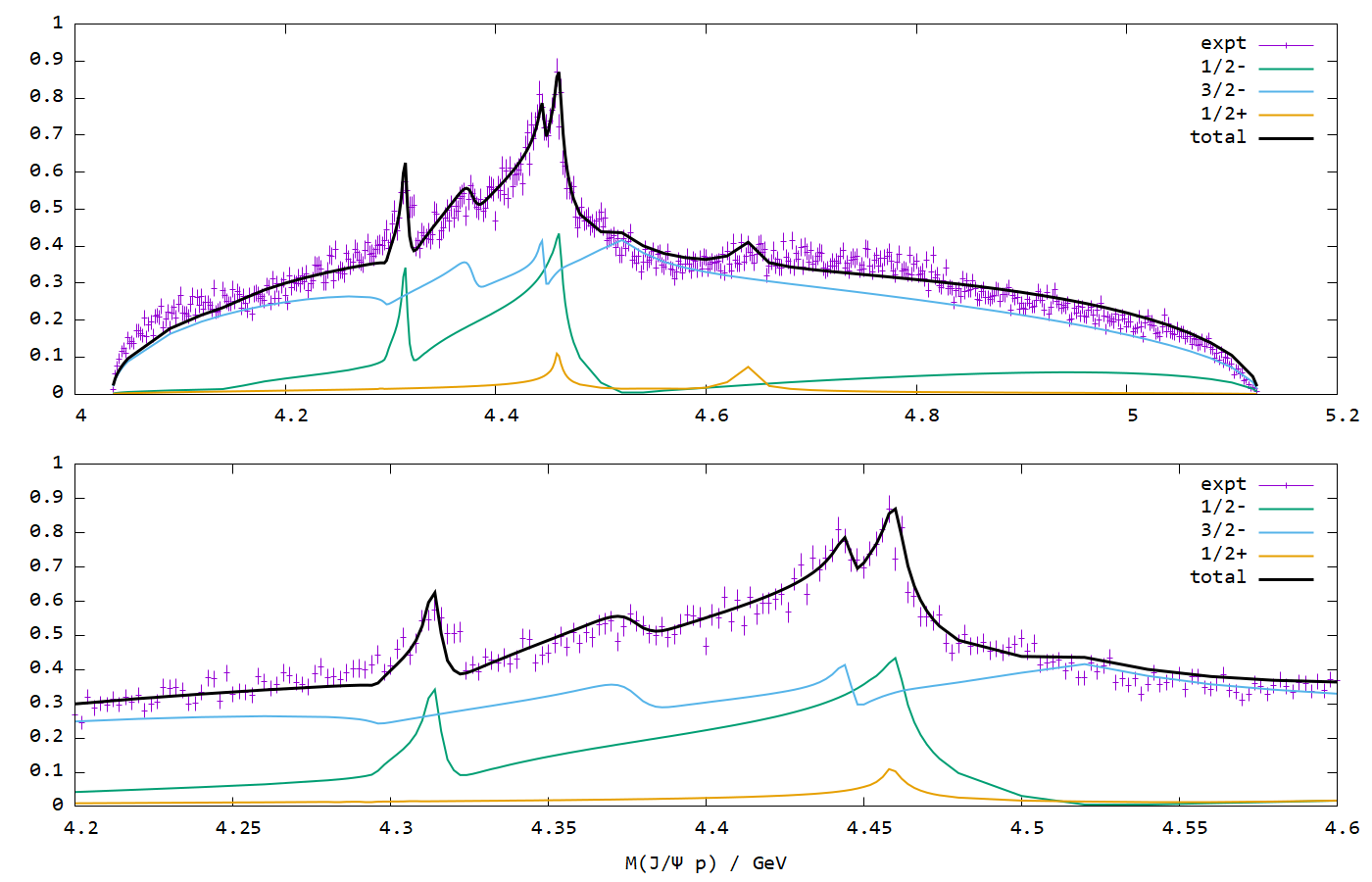}
    \caption{The $\jpp$ invariant mass spectrum in Case~5, where the $1/2^+$ peak at $P_c(4457)$ is due to a $\Lc(2595)\D$ resonance.}
    \label{fig:fit5}
\end{figure}

In  Fig.~\ref{fig:fit5} we give an example fit, with $F_a = -12$ GeV$^{-2}$. The quality of the fit indicates that the $P_c(4457)$ peak is consistent with the presence of a resonance at this mass, however when comparing to our previous fits, we would not argue that the data require one. Because the model is an elaboration of previous cases, we therefore prefer their relative parsimony of description.

Of course with $F_b=0$, there is identical attraction in all three channels $\Lc(2595)\D$, $\Lc(2595)\D^*$ and $\Lc(2625)\D^*$ (see Table~\ref{tab:3}), so a resonance in $\Lc(2595)\D$ would imply partner states in $\Lc(2595)\D^*$ and $\Lc(2625)\D^*$. The heavier state $\Lc(2625)\D^*$ contributes something useful to the fit, helping to account for the rising feature in the data above 4.6~GeV. The other state $\Lc(2595)\D^*$ is invisible in the fit, as there is no prominent feature near the corresponding threshold, so the fit results in a small production coupling (Table~\ref{tab:fits}).

In any case, the existence of these partner states should not be interpreted as a robust prediction of the resonance interpretation of $P_c(4457)$, since it follows from our choice of contact terms which, as noted, is not well constrained. For example, we could presumably remove the additional states from the spectrum by making $F_b$ large and negative, and re-tuning $F_a$ to compensate the increased off-diagonal attraction in the $\Lc(2595)\D$ channel. In this case, the feature at $\Lc(2625)\D^*$ threshold would be linked to the triangle cusp rather than a corresponding resonance. 

\subsection{Discussion of Fit Results}
\label{sect:disc}

Our studies convincingly demonstrate that the LHCb states $P_c(4312)$, $P_c(4380)$, and $P_c(4440)$ are associated with strong final state interactions in $\Sigma_c\bar{D}$, $\Sigma_c^*\bar{D}$, and $\Sigma_c\bar{D}^*$ respectively. Here we seek to characterize these interactions more fully by investigating the analytic structure of the final state T-matrix in more detail. This is done with a simple search for poles in the complex energy plane on the closest physical sheet (defined by the negative (positive) square root of the breakup momentum for all channels with thresholds below (above) the real part of the pole). Residues to various decay channels are  obtained with the aid of a Cauchy integral centred on the pole in the appropriate sheet for the diagonal T-matrix element in the channel of interest. These residues are then multiplied by the phase space (at the pole position) to give the partial widths reported in Table \ref{tab:poles}. 

The T-matrix is computed by solving the Bethe-Heitler equation, which amounts to performing a sum over bubble diagrams, some of which contain particle propagators with finite widths. These widths shift pole locations by approximately $\Gamma/2$, where $\Gamma$ is the relevant resonance total width. In this system the resonances with substantial widths are the $\Sigma_c^*$ (15~MeV) and the $\Sigma_c$ (1.86~MeV), both of which decay predominantly to $\Lambda_c\pi$. In the case of weakly bound systems, such  as those considered here, constituents approximately contribute their widths to the full width of the bound state (a formalism for describing this in more detail can be found in Appendix A of Ref.~\cite{Swanson:2006st}), depending on how much the particular resonance channel contributes to the bound state. In our case, the $P_c(4312)$ is dominated by $\Sigma_c\bar{D}$, the $P_c(4380)$ is dominated by $\Sigma_c^*\bar{D}$, and the $P_c(4440)$ is dominated by $\Sigma_c\bar{D}^*$. Thus we expect contributions of approximately 1.9, 15, and 1.9 MeV to the respective total widths~\cite{Burns:2021jlu}. These are reported in Tab.~\ref{tab:poles} in the rows labelled ``three body". The sum of the partial widths should  approximately equal twice the imaginary part of the pole position, and we find that this is indeed the case, supporting the assumptions made.

\begin{table}[ht]

\begin{tabular}{l|cccc}
\hline\hline
     & Case~2a/3 & Case~2b & Case~4 & Case~5 \\
\hline
$1/2^-$ pole  & $4312 -2.4i$ & $4308 - 5.2i$ & $4312 - 2.6i$ &  $4312 - 2.4i$ \\
\phantom{$1/2^-$} $\Lambda_c\D$ & 0.0086   & 0.0194   &  0.0080&  0.0086 \\
\phantom{$1/2^-$} $\Lambda_c\D^*$ & 3.10  & 9.8   &  3.4 &  3.1 \\
\phantom{$1/2^-$} $N J/\psi$ & 0.0015  & 0.0011    &  0.0017 &  0.0015 \\
\phantom{$1/2^-$} $N \eta_c$  & 0.0055  & 0.0041    & 0.0060 &  0.0055 \\
\phantom{$1/2^-$}three body  & 1.86  &  1.86   & 1.86  &  1.86 \\
\hline
$3/2^-$ pole & $4376.5 - 9.0i$  &  $4375 - 10.7i$ & $4376.5 -9.1i$ & $4376.5 - 8.9i$\\
\phantom{$3/2^-$} $\Lambda_c\D^*$  & 3.11 & 7.2  & 3.41 & 3.12 \\
\phantom{$3/2^-$} $N J/\psi$  & 0.0088 & 0.0085  & 0.0097 & 0.0088 \\
\phantom{$3/2^-$}three body  & 15  &  15   & 15  & 15  \\
\hline
$3/2^-$ pole  & $4444 - 2.56i$  &  $4440 - 3.3i$ & $ 4444.5 - 2.56i$ & $4444 - 2.56i$\\
\phantom{$3/2^-$} $\Lambda_c\D^*$ & 1.16 & 2.2  & 1.25 & 1.16 \\
\phantom{$3/2^-$} $\Sigma_c^*\D$  & 1.7 & 2.1  & 1.58 & 1.70 \\
\phantom{$3/2^-$} $N J/\psi$ & 0.0041 & 0.0036  & 0.0045 & 0.0041 \\
\phantom{$3/2^-$}three body & 1.86  &  1.86   & 1.86  & 1.86  \\
\hline
$1/2^+$ pole  &   &   &  & $4458 - 0i$\\
\hline\hline
\end{tabular}
\caption{Pole positions and partial widths (MeV).}
\label{tab:poles}\end{table}

There is a compelling and simple pattern in the numbers in Table~\ref{tab:poles}. The widths of the $P_c(4312)$, $P_c(4380)$ and $P_c(4440)$ are dominated by  $\Lc\D^*$,   $\S^*\D$, and three-body decays. By comparison, the decays to $\Lc\D$, $N\jp$ and $N\eta_c$ are tiny, which is consistent with the upper limits~\cite{Burns:2021jlu} implied by experimental data. For $P_c(4312)$, the significant suppression of $\Lc\D$ relative to $\Lc\D^*$ is due to a selection rule which arises from heavy quark symmetry~\cite{Voloshin:2019aut}, and which is essential in explaining the upper limits on $P_c$ fit fractions in  $\Lambda_b\to \Lc\D^{0}K^-$ decays~\cite{Burns:2021jlu}.

We have not been able to identify the width of the resonance corresponding to $P_c(4457)$ in Case~5, for numerical reasons. It is extremely narrow, and is manifest as a large enhancement near 4458 MeV that is very close to the real axis. Increasing the magnitude of $F_a$ merely shifts this location to lower energy. The narrow width is consistent with the coupled channel model used (see Table \ref{tab:3}) since the only permitted decay modes in the model are very weakly coupled. 

Experimental total widths are determined to be
\begin{align}
\Gamma[P_c(4312)]&=9.8\pm 2.7^{+3.7}_{-4.5}~\textrm{MeV},\label{eq:G1}\\
\Gamma[P_c(4380)] &= 208 \pm 18 \pm 86~\textrm{MeV}, \\
\Gamma[P_c(4440)]&=20.6\pm4.9^{+8.7}_{-10.1}~\textrm{MeV},\label{eq:G2}\\
\Gamma[P_c(4457)]&=6.4\pm 2.0^{+5.7}_{-1.9}~\textrm{MeV},\label{eq:G3}
\end{align}
whereas our estimates for the total widths are 
\begin{align}
\Gamma[P_c(4312)]&= 5 - 8~\textrm{MeV},\\
\Gamma[P_c(4380)] &\approx 18 - 22~\textrm{MeV},\\
\Gamma[P_c(4440)]&= 5 - 6~\textrm{MeV}.
\end{align}

The predicted width of the $P_c(4312)$ is reasonably close to that measured, while that of $P_c(4440)$ is somewhat too small. There is a much larger discrepancy in the case of $P_c(4380)$, although we remark that the evidence for this state comes from the amplitude analysis in the first LHCb results in this system~\cite{Aaij:2015tga}, which has been superseded by the more recent results of Ref.~\cite{Aaij:2019vzc}, which render the measured properties of $P_c(4380)$ obsolete. In the molecular scenario, the mass of $P_c(4380)$ is necessarily near $\Sigma_c^*\bar{D}$ threshold, and its width is constrained by heavy-quark symmetry to be comparable to those of the other $P_c$ states, though somewhat larger because of the intrinsic width of the $\S^*$ constituent. For this reason, molecular models generically predict a $P_c(4380)$ width which is considerably smaller than the measured value (see for example, refs~\cite{Du:2019pij,Du:2021fmf}). We therefore suggest that it is too early to consider the lack of agreement as negative evidence, and recommend that narrow states near 4380 MeV be considered when constructing future amplitude models.

It is noteworthy that our fits very nicely reproduce the lineshapes of the $P_c$ states, yet the resonance widths (as extracted from the T-matrix) are not entirely consistent with those measured in experiment -- not just for the $P_c(4380)$, which is exceptional for the reasons outlined above, but also for $P_c(4440)$. This underlines the point that the lineshape in our approach is a function not only of the resonance properties (mass and width), but also the interference between the resonance and other contributions (in our case, from the triangle diagrams and the constant background). Consequently, the width of the lineshape is not necessarily the same as the intrinsic width of the resonance. This is quite different to the experimental analysis, in which the measured width of the lineshape (having subtracted off an incoherent background) is attributed to a resonance only. Hence it unsurprising that our widths are not identical to those of experiment. Conversely, the extracted widths of Du~\etal\cite{Du:2019pij,Du:2021fmf} are somewhat closer to those of experiment than our own, and this is likely because their procedure is effectively very similar to that of experiment~\cite{Aaij:2019vzc}: by combining signal and background incoherently, the width of the lineshape is effectively the intrinsic width of the resonance.

Residue values are also of interest since they indicate the strength of resonance couplings to
various channels. We report residues for all channels for Case 2a in Table \ref{tab:residues}. These results indicate that couplings to the $J/\psi N$, $\eta_c N$, and $\Lambda_c \bar{D}$ channels are all very weak, 
consistent with the small branching fractions reported above. Couplings to $\Lc\D^*$ and $\Sigma_c^{(*)}\bar{D}^{(*)}$ channels are stronger, with the largest coupling being in the $\Sigma_c^{(*)}\bar{D}^{(*)}$ channel with the nearest threshold
-- again, as expected.  

Our production model (Sect.~\ref{sec:production}) attributes the dominance of diagram (a) to the enhanced electroweak vertex, and the claim that the final state interactions (what we call the ``$P_c$ vertex'' in Fig.~\ref{fig:production}) in diagrams (a) and~(c) are comparable.
The latter assumption may appear to be in contradiction with the reported residues, which indicate the dominance of $\Sigma_c^{(*)}\bar{D}^{(*)}$ components in the $P_c$ resonances, possibly suggesting an enhancement of the final state interactions in diagram (c). In fact this is not the case. The
final state interactions that generate $\Sigma_c^{(*)}\bar{D}^{(*)}$ bound states are common to diagrams (a) and~(c) and therefore enhance both approximately equally.  
What distinguishes the two diagrams is the couplings of the triangle legs to the dominant $\S\*\D\*$ components in the resonances, namely the relative sizes of the couplings $\Lc\D\*\to\S\*\D\*$ (diagram (a)) and $\S\*\D\*\to\S\*\D\*$ (diagram (c)). We previously argued that these couplings are comparable, and these arguments are confirmed by our fitted values of $B$, $C_a$ and $C_b$.

We have taken the precaution of computing diagram (c) and confirm that, 
before taking account of the suppression at the electroweak vertex, it is comparable to diagram (a).

\begin{table}[ht]
\begin{tabular}{l|c|c|c}
\hline\hline
channel & $1/2^- (4312.0-2.4375i)$ & $3/2^- (4376.5-8.88i)$ & $3/2^- (4444.0-2.56i)$ \\
\hline
$\Lambda_c\bar{D}$ & $(11.17-1.28i)\cdot 10^{-5}$ &  &  \\
$\Lambda_c\bar{D}^*$ & $(4.91+1.45i)\cdot 10^{-2}$ & $(3.30+1.42i)\cdot 10^{-2}$ & $(13.89-1.38i)\cdot 10^{-3}$ \\
$\Sigma_c\bar{D}$ & $0.888+0.162i$ &  &  \\
$\Sigma_c^*\bar{D}$ &  & $0.865+0.151i$ & $(-1.012+1.74i)\cdot 10^{-2}$ \\
$\Sigma_c\bar{D}^*$ & $0.186-0.0377i$ & $0.108-0.035i$ & $1.74 + 0.172i$ \\
$\Sigma_c^*\bar{D}^*$ & $(9.71-2.25i)\cdot 10^{-2}$ & $0.285-0.054i$ & $0.248 -0.151i$ \\
$J/\psi N$ & $(27.1-8.87i)\cdot 10^{-6}$ & $(16.8-5.44i)\cdot 10^{-5}$ & $(6.27-5.89i)\cdot 10^{-5}$ \\
$\eta_c N$ & $(10.9-4.42i)\cdot 10^{-5}$  &  &  \\
\hline\hline
\end{tabular}
\caption{Residues for Case 2a (GeV${}^{-1}$).}
\label{tab:residues}
\end{table}

\section{Discussion and Conclusions}
\label{sec:conclusions}

We have argued that a complete understanding of the reaction $\Lambda_b \to \jpp\, K^-$ requires $\Lambda_c^{(*)}\bar{D}^{(*)}$ degrees of freedom. This claim is supported by the data itself, electroweak phenomenology, simple theoretical arguments, and recent  measurements of $\Lambda_b$ decay modes. It is also clear that the closely related $\Sigma_c^{(*)}\bar{D}^{(*)}$ system plays a significant role. With these considerations, it is natural to model the reaction with a triangle diagram that contains an electroweak vertex at one apex and strong interactions at the other apexes. Thus the $\jpp$ subsystem emerges from strong final state interactions coupled to a production triangle subgraph.

We have constructed a model that incorporates the known experimental constraints concerning $\Lambda_b$ decay modes, that respects well-established electroweak phenomenology and heavy quark symmetry, and is consistent with one-pion exchange dynamics. The model is capable of fitting the entire mass spectrum,  and does not require unnatural ``explanations" for missing $\Sigma_c^*\bar{D}^*$ states.

As well as the production mechanism, another distinguishing feature of our model is that, unlike many alternative models, we do not assume that $P_c(4457)$ is a $\S\D^*$ bound state. Instead we find that it can be described as a $\S\D^*$ threshold cusp, an enhancement due to the $\Lc(2595)\D$ triangle singularity, or a $\Lc(2595)\D$ resonance. 

A simple version of the model (Case~1) reveals that 
many of the features of the $J/\psi\, p$ spectrum can be explained in terms of a constant background and the postulated production triangle amplitude. In particular, the $P_c(4457)$ emerges as a $\S\D^*\to\jpp$ threshold cusp, similar to the mechanism advocated in ref.~\cite{Kuang:2020bnk}.

Enabling final state interactions that are capable of forming bound states in the $\Sigma_c^{(*)}\bar{D}^{(*)}$ system, we get an improved fit (Case~2), with sharp resonance peaks for established states $P_c(4312)$ and $P_c(4440)$, and a broader peak corresponding to the enhancement previously identified as $P_c(4380)$. The broader width of $P_c(4380)$ is a natural consequence of the intrinsic widths of its dominant constituents, while the widths of the other states are primarily controlled by $B$, which is fit to data and which takes a value that, in comparison to $C_b$, is roughly consistent with expectations from one-pion exchange.

Introducing the $1/2^+$ channel brings further improvements in the fit in the higher mass range, owing to production via $\Lc(2595)\D\*$ and $\Lc(2625)\D^*$ diagrams. In a minimal extension of the previous cases, we find a substantial improvement in the fit only if we include $1/2^+$ background (Case~3), which is less satisfactory phenomenologically, as it implies the spectrum is dominated by $\jpp$ in P-wave. We get a  better result (Case~4) by tuning the ``$D_s$'' mass to reveal the logarithmic singularity in the $1/2^+$ production triangle diagrams, resulting in a sharp $P_c(4457)$ peak, and no need for background in the $1/2^+$ channel. A fit of comparable quality can also be obtained by introducing attractive interactions in the  $1/2^+$ channel (Case~5), resulting in a $\Lc(2595)\bar{D}$ resonance at 4457 MeV, however this step is clearly not required given the quality of the previous models. 

Given the comprehensive agreement with a range of experimental and theoretical constraints, we believe that these results constitute firm evidence for novel meson-baryon resonances and for the importance of ``kinematical" effects such as created by triangle diagrams in certain reactions.

Our results reinforce our previous arguments that $P_c(4457)$ should not be considered as a $\S\D^*$ bound state partner to $P_c(4440)$. We already pointed out \cite{Burns:2021jlu} that the widespread assumption (Scenarios~A and B) that both $P_c(4440)$ and $P_c(4457)$ are $\S\D^*$  bound states is problematic phenomenologically, as it contradicts heavy quark symmetry relations between $\Lc\D$ and $\Lc\D^*$ decays, and implies several $\S^*\D^*$ partners which are apparently not seen in experiment. We showed that we can avoid these problems by assuming that only $P_c(4440)$ is a $\S\D^*$ bound state, with $3/2^-$ quantum numbers (Scenario~C). To complete the picture, we need an alternative explanation for $P_c(4457)$, and we argued in our previous paper that there are several plausible alternatives, all arising naturally from the proximity of the state to $\S\D^*$ and $\Lc(2595)\D$ thresholds. We have now verified that all of these alternative scenarios can reproduce experimental data.

Indeed our current results provide additional arguments in favour of Scenario~C. One of the issues with Scenarios~A and B is that they imply $\Sigma_c^*\bar{D}^*$ partners ($1/2^-$, $3/2^-$, and $5/2^-$) which are are conspicuously absent from the data. In other models~\cite{Du:2019pij,Du:2021fmf}, one has to assume (without explanation) that the production of these states is small, exploiting the intrinsic parametric freedom in the production model. In our model we do not have such freedom, and we find on the contrary that the production of the missing $3/2^-$ state, in particular, is not small, but very much enhanced compared to that of other states. Hence it is not possible to explain this state away. This follows straightforwardly from heavy-quark symmetry: the relative rates of the different bound states can be estimated from the square of the product of the production and decay matrix elements, namely the entries proportional to $B$ and $E$ respectively. In the $3/2^-$ channel, this algebra implies that the missing $\S^*\D^*$ state is enhanced by a factor of 25 compared to the $\S\D^*$ state, so if it binds, its resonance peak would be enormous in comparison to $P_c(4440/4457)$. This is of course strikingly inconsistent with data. Indeed we have verified that it is not possible, with our model, to obtain a good fit in which both $P_c(4440)$ and $P_c(4457)$ are $\S\D^*$ bound states, for precisely this reason. 

In our Scenario C, the experimental absence of prominent $\S^*\D^*$ states is not a problem. The $1/2^-$ and $3/2^-$ states simply do not bind -- this is a natural feature of the parameter space relevant to Scenario~C. The $5/2^-$ state does bind, but its absence in experiment is quite natural. As noted elsewhere~\cite{Xiao:2013yca,Takeuchi:2016ejt,Shimizu:2019ptd,Xiao:2019aya,Burns:2019iih,Burns:2021jlu}, the $\jpp$ decay of this state is D-wave, and so is naturally suppressed compared to the decays of other $P_c$ states, which are S-wave. In our model, there is a further suppression due to the production mechanism: the $5/2^-$ state couples to the assumed production channels ($\Lc\D$ and $\Lc\D^*$) in D-wave, whereas the other $P_c$ states couple in S-wave. 

The successful phenomenology associated with the suppression of D-wave terms gives some justification for our explicit assumption, from the outset, that the production and decay of $P_c$ states are dominated by S-wave interactions. (We make an exception in the case of the $1/2^+$ channel, for which $\jpp$ is P-wave.) In this respect our model differs from that of Du \etal\cite{Du:2019pij,Du:2021fmf}, where D-wave terms are included and are found to be significant. We find no need to include D-wave terms, having obtained excellent fits with S-wave interactions alone. Furthermore, we would argue that models in which the D-wave contributions are large are less satisfactory phenomenologically, as they do not have a simple explanation for the absence of the $5/2^-$ state. Such models also have more parameters (contact terms).

Our model has a number of generic features that arise naturally from heavy quark symmetry, all of which compare favourably with experimental data. In order to generate a (dominantly) $\S\D$ state describing $P_c(4312)$, we automatically also have a $\S^*\D$  partner state near 4380 MeV, as (from heavy quark symmetry) the two channels have the same diagonal potentials. This feature is clearly present in the data, and moreover its larger apparent width can be understood as due to the intrinsic width of its constituents. 

The $P_c(4440)$ is considerably more bound than $P_c(4312)$, and to achieve this we have to introduce non-zero $C_b$, to make its potential more attractive. From the heavy quark symmetry (the pattern of binding potentials), this automatically implies a heirarchy of binding potentials among the $\S\D^*$ and $\S^*\D^*$ states. The resulting pattern is particularly favourable phenomenologically if we choose $C_b<0$, because it implies that,  of the $\S^*\D^*$ states, only $5/2^-$ binds (not $1/2^-$ or $3/2^-$), which is  good for the reasons outlined above. The same choice ($C_b<0$) also works nicely because it implies that, of the $\S\D^*$ states, only $3/2^-$ binds (not $1/2^-$), explaining why the $P_c(4440)$ apparently decays sparingly to $\Lc\D$ (as it couples only in D-wave) \cite{Burns:2021jlu}.  The choice $C_b<0$ is also consistent with the pattern of binding due to one-pion exchange.

Another nice feature of heavy quark symmetry is that it explains the tight upper limit on $P_c(4312)\to\Lc\D$ decay \cite{Burns:2021jlu}. Despite coupling in S-wave, the transition is forbidden by heavy quark symmetry, assuming the dominance of the $\S\D$ component \cite{Voloshin:2019aut}.

Searches for the $P_c$ states in photoproduction have given null results that impose constraints on our model. Specifically, the heavy quark parameter $E$ controls the coupling of the $\Sigma_c^{(*)}\bar{D}^{(*)}$ to $J/\psi N$. We have shown that the ratio $|B/E|$ is fixed by the $P_c(4312)$ $\Lambda_c\bar{D}^*$ and $\jpp$ branching ratio fraction. Since $B$ is related to pentaquark widths we are then able to obtain a reasonably reliable estimate for $E$. Finally, we obtain $J/\psi N$ partial widths of

\begin{align}
&P_c(4312): \ \  1-2 \ \textrm{eV} \nonumber \\
&P_c(4380): \ \  9-10 \ \textrm{eV} \nonumber \\
&P_c(4440): \ \ 4-5 \ \textrm{eV}. \nonumber \\
\end{align}
These comprise branching fractions of approximately $10^{-3}$ in all cases, which implies that the current experiments are at the threshold of being able to observe the $P_c$ states. We therefore encourage continued effort in photoproduction searches. 

We now offer some predictions, and suggestions for future experimental analyses which can test aspects of our model, or discriminate among the different cases we have investigated.

Clearly, measuring the quantum numbers of the $P_c$ states will be helpful in supporting the molecular hypothesis, and in discriminating among competing molecular scenarios. Models typically agree on the quantum numbers of $P_c(4312)$ and $P_c(4380)$, namely $1/2^-$ and $3/2^-$, respectively, and these predictions can be tested in experiment. The quantum numbers of $P_c(4440)$ and $P_c(4457)$ are more discriminating: a robust prediction of our preferred Scenario~C is that $P_c(4440)$ has $3/2^-$ quantum numbers. Discerning the quantum numbers of the $P_c(4457)$ is likely to be experimentally more challenging since, as is evident in several cases we consider, the signal can be a superposition of contributions from different quantum number channels. Measurement of quantum numbers requires an amplitude analysis, and reliable results require all relevant states to be included. With this in mind, we advocate that a resonance near 4380~MeV be built into future amplitude models, as this is a generic feature of molecular models.

Resonances can in principle be distinguished from other effects (cusps, triangle singularities) by studies in other production and decay modes. In our preferred scenario, $P_c(4312)$, $P_c(4380)$, and $P_c(4440)$ are all resonances. An indication of their resonant nature would be their observation in other production modes (such a photoproduction), and other decay channels (such as $\eta_c N$, $\Lc\D\*$ and $\S\*\D$), and in particular, finding that their measured masses and widths are the same in these various processes. (The caveat here is that their apparent widths, as obtained for example by fitting an incoherent background and a Breit-Wigner peak, may differ, for the reasons alluded to previously.)

The same is not necessarily true of $P_c(4457)$, so its existence or otherwise in various production and decay modes could be revealing. For example, as a resonant state (Case~5) it would, like the other states, presumably be visible in other production and decay modes, and with the same measured properties. However, if the $P_c(4457)$ were due to the logarithmic singularity in the triangle diagram (Case~4), it would presumably be completely absent in, for example, photoproduction, because the corresponding triangle appears only in a convoluted Feynman diagram which we expect makes negligible contribution to the cross section.

The role of the triangle singularity for $P_c(4457)$ can also be tested in other ways. For example, it would imply a dip in the $\Lambda_b\to \Lc(2595)\D^0 K^-$ spectrum for $\Lc(2595)\D^0$ invariant mass around 4457~MeV, as noted in ref.~\cite{Aaij:2019vzc}, making use of the analysis of ref.~\cite{Szczepaniak:2015hya}. Of course it also implies a $D_s$ state around 2.92~GeV, which remains to be established experimentally. Additional diagnostics of triangle singularities are discussed in ref.~\cite{Guo:2019twa}.

In our previous paper, we gave a host of predictions for branching fraction and fit fractions of $P_c(4312)$, $P_c(4380)$ and $P_c(4440)$, in $\Lambda_b\to\jpp\, K^-$, $\Lambda_b\to\eta_c\, p\, K^-$, $\Lambda_b\to \Lc\D^{0(*)}K^-$ and $\Lambda_b\to \S\*\D K^-$, all of which can be used to test our preferred Scenario~C. Within our current model, we can in principle go further, and make predictions not only for the branching and fit fractions, but also the invariant mass spectra. However in practice, detailed predictions are difficult. For example, in $\Lambda_b\to\jpp\, K^-$, the tree-level contribution is colour-suppressed, and is considered as a background which is fit to data: we are effectively assuming that the interesting features in the $\jpp\, K^-$ spectrum arise through colour-favoured triangle diagrams with intermediate $\Lc\D\*$ states. The situation is very different in $\Lambda_b\to \Lc\D^{0(*)}K^-$, where the same final state arises through both tree-level and triangle diagrams, each of which is colour-favoured, and where the tree-level contributions will contribute prominent features due to $D_s$ resonances, which are well beyond the scope of our model. The $\Lambda_b\to \S\*\D K^-$ decay is somewhat less complicated, because of the suppression of the tree-level contribution, but still the impossibility of predicting the three-body background makes reliable predictions difficult.

With these complications in mind, the safest prediction that we can provide is for $\Lambda_b \to \eta_c p K^-$, because the  couplings of the $P_c$ states to $\eta_c p$ are fixed by heavy quark symmetry, while the reaction dynamics are nearly identical to those of $\Lambda_b \to \jpp\, K^-$. We start with some general remarks on what to expect. The $\eta_c p$ channel is a potentially interesting filter on $P_c$ quantum numbers, since in S-wave it couples to $1/2^-$ only, whereas $\jp p$ couples to both $1/2^-$ and $3/2^-$. Comparison of the two spectra can thus give some indication of the likely quantum numbers of the states. In our Case~2, for example, we may expect the $\eta_c p$ distribution to show a resonance peak at $P_c(4312)$, and a cusp at $P_c(4457)$, but no structure corresponding to $P_c(4380)$, $P_c(4440)$, or at the higher threshold $\S^*\D^*$. We can make these predictions more precise by drawing on heavy quark symmetry, which gives some indication of how the $1/2^-$ structures in $\eta_c p$ compare to the corresponding features in $\jp p$. With reference to the matrix elements in Table~\ref{tab:1}, there is an enhancement by a factor of 3 in $P_c(4312)$, but a suppression of 3/25 in $P_c(4457)$ \cite{Sakai:2019qph,Voloshin:2019aut,Burns:2021jlu}. Hence we expect the $\eta_c p$ spectrum to exhibit a prominent $P_c(4312)$ peak, but less evidence for $P_c(4457)$.

\begin{figure}
    \centering
    \includegraphics[width=\textwidth]{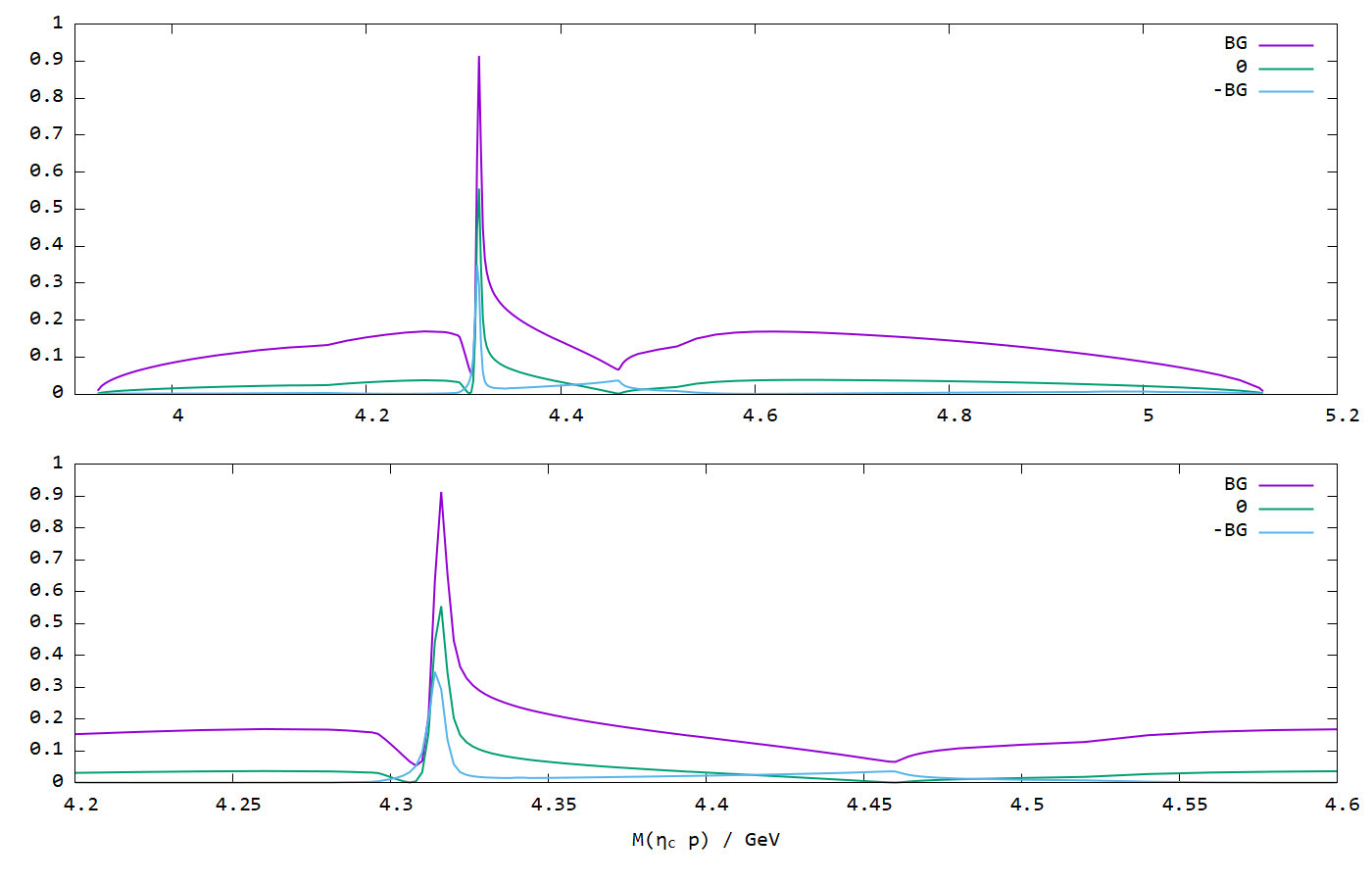}
    \caption{Prediction for the $\eta_c p$ invariant mass spectrum in $\Lambda_b \to \eta_c p K^-$ decays, with the production couplings and contact terms corresponding to Case~2a (Table~\ref{tab:fits}). The three curves correspond to different models for the background: fixed to that of $J/\psi N$ (``BG''), set to zero (``0''), and with opposite sign to $\jp N$ (``$-$BG'').}
    \label{fig:111}
\end{figure}

The precise shape of the spectrum is very difficult to predict, because it depends on $\eta_c \,p$ background, which may be quite different to $\jpp$.  We therefore provide some illustrative examples (Fig.~\ref{fig:111}) of the $\eta_c \,p$ distribution (in Case~2a), taking three different models for the background: fixed to that of $J/\psi\, p$, set to zero, and with opposite sign to $\jpp$. As anticipated above, the $P_c(4312)$ resonance forms a prominent peak in all cases, but the $P_c(4457)$ cusp is hardly visible. (Notice that the cusp can manifest as a peak or dip, depending on the background.)

The suppression of $P_c(4457)$ in $\eta_c \,p$ is a particular feature of Case~2, which is not necessarily true of cases 3, 4 or 5. In the latter cases, the $P_c(4457)$ peak has a contribution from the $1/2^+$ channel $\Lc(2595)\D$, either as a triangle cusp, an enhancement due to the triangle singularity, or a resonance. Because this channel couples to $\jpp$ and $\eta_c\,p$ in the same partial wave, we may expect $P_c(4457)$ signals in both $\jpp$ and $\eta_c\,p$. This is quite different to Case~2 in which the $P_c(4457)$ is hardly visible in $\eta_c \,p$. Hence if the $\eta_c \,p$ spectrum shows a prominent $P_c(4457)$ peak, it is an indication of the role of the $1/2^+$ channel.

Referring to Table~\ref{tab:3}, the relative rate of $\jpp$ and $\eta_c\,p$ in $1/2^+$ follows from heavy-quark symmetry, and depends on the parameters $G_a$ and $G_b$. Ignoring phase space differences, the ratio of rates is
\begin{align}
    \frac{R(\jpp)}{R(\eta_c\,p)}=\frac{1}{3}+\frac{8}{3}\left(\frac{G_b}{G_a}\right)^2
\end{align}
In principle, the relative rates of $\jpp$ and $\eta_c\,p$ could be used to fix $G_b/G_a$, which is currently poorly constrained by data. In practice this is not really possible, since the formula applies only to the $1/2^+$ channel, whereas in our model, the $P_c(4457)$ peak is a superposition of this and other contributions (primarily $1/2^-$). 

A solid prediction of our Scenario~C is the existence of a $5/2^-$ $\S^*\D^*$ partner state. We have argued that this state is not prominent in $\Lambda_b\to\jpp\, K^-$ because of suppressed production and decay, however it would be interesting to include such a state in future amplitude analyses with higher statistics. The state could also be revealed in other production and decay modes. An intriguing possibility is the decay $\jp\Delta$, which may arise for this and other $P_c$ states because of the anticipated isospin mixing in their wavefunctions \cite{Burns:2015dwa}. Although the branching fraction is expected to be small, and the experimental analysis will be difficult, an advantage is that the $5/2^-$ state decays to $\jp\Delta$ in S-wave, and so should at least be comparable to the same decays for the other $P_c$ states. This is quite different to the case of $\jpp$, where the $5/2^-$ state is suppressed compared to the other $P_c$ states.

There are in addition some experimental measurements which can test our proposed production mechanism. With reference to Fig.~\ref{fig:production}, a crucial assumption is that the electroweak vertex in diagram (c) is suppressed in comparison to that of diagram (a). (The suppression of diagram (b) is already established experimentally, and is anyway less relevant because of the additional suppression at the strong vertex.) As a test of the suppression of diagram (c), we urge the experimental measurement of $\Lambda_b\to \Xi_c\*\D\*$, anticipating significant suppression in comparison to the (already measured) $\Lc^+\D_s^{-(*)}$ modes. Similarly, we suggest measurement of $\Lambda_b\to\S\*\D\* K^-$, expecting this to be small in comparison to the measured $\Lambda_b\to\Lc\D^{0(*)}K^-$.

We conclude with some observations about the importance of combining amplitudes coherently where relevant, something which is of course very well-known but which, for understandable reasons of pragmatism, is often ignored in experimental or theoretical analyses. It is commonplace in both experiment and theory analyses (for example refs.~\cite{,Aaij:2019vzc,Du:2019pij,Du:2021fmf}) to fit data through an incoherent combination of a background (modelled in some way) and signal (simply fit as a Breit-Wigner distribution, for example, or derived from non-perturbative interactions). In this approach, separating signal from background is a matter of taste, influenced heavily by the level of complexity one is willing to tolerate in the background model. The end result is that any features which cannot be subsumed into a smooth background are defined as signal, often interpreted (implicitly or explicitly) as hadronic resonances. But the outcome can be totally different if the background and signal are combined coherently, as exemplified by our simplest model (Case 1). We find that the simplest possible background (a complex constant), combined coherently with a signal amplitude which has cusps at several thresholds, can reproduce the overall shape of the spectrum -- including the sharp $P_c(4457)$ peak -- with no need for any resonances. It works in this particular case because the sharp feature is associated with a threshold ($\S\D^*$), where the amplitude naturally has a cusp; we do not expect a similar mechanism to account for any sharp  feature away from threshold. Hence we emphasise the need for caution in the interpretation of features which are close to thresholds.

Interference effects can also confound the extraction of resonance properties, particularly their widths. Again, this point is very well-known in the literature, but we mention it here again because its significance is particularly apparent in our results. As noted previously, we obtain good fits to the resonant peaks for all $P_c$ states, although the widths we extract from the T-matrix are smaller than the values measured in experiment. The difference is because the lineshape in our approach arises from interfering amplitudes with contributions from the resonance, as well as the triangle and background terms. Unsurprisingly, the width of the resonant contribution in this approach is not necessarily consistent with that obtained in experiment \cite{Aaij:2019vzc}, where the lineshape is attributed to a Breit-Wigner resonance combined incoherently with the background. (Note that the experimental analysis did consider the effect of interference among different $P_c$ states with the same quantum numbers, as a means of estimating systematic uncertainties;  the interference effects in our model are more pronounced and, being more model-dependent, cannot easily be included in experimental analyses.)

In summary, we have argued for the importance of $\Sigma_c^{(*)}\bar D^{(*)}$ and $\Lambda_c^{(*)}\bar D^{(*)}$ channels combined with ``kinematical" effects in describing the LHCb $P_c$ signals. Our resulting model is consistent with experimental constraints, heavy quark symmetry, and electroweak phenomenology and provides strong evidence for novel meson-baryon bound states and sharp non-resonant effects in hadronic systems.

\acknowledgments
Swanson's research was supported by the U.S. Department of Energy under contract DE-SC0019232.

\bibliography{bibfile}

\appendix
\section{Iterated Triangle Diagrams}

We have stressed the role that final state interactions in the $\Lambda_c \bar{D}-\Sigma_c\bar{D}$ system plays in creating the LHCb $P_c$ signals.  It is also possible to generate final state interactions by iterating the production triangle diagrams of Fig.~\ref{fig:production}\footnote{We thank M. Hansen for comments that led to these considerations.}. Here we show that these diagrams make a small contribution to the reaction.

\begin{figure}
    \centering
    \includegraphics[width=\textwidth]{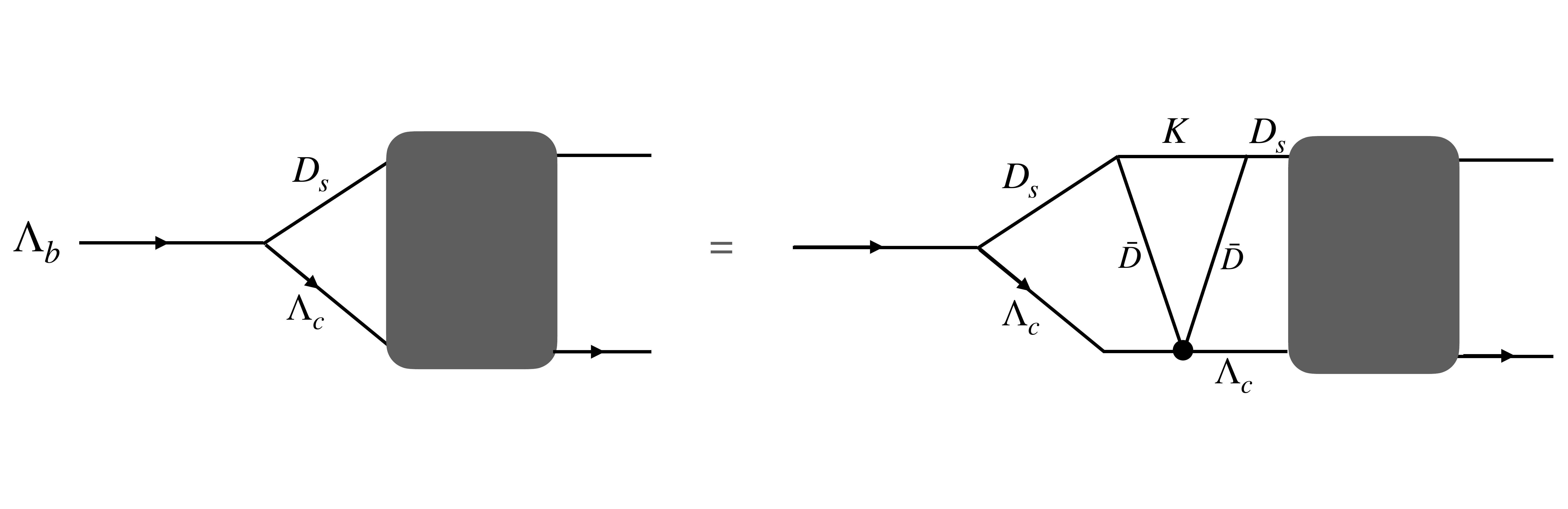}
    \caption{Iterating Triangle Diagrams }
    \label{fig:FSI-tri}
\end{figure}

The triangle diagrams considered here can be iterated to form a final state interaction as illustrated in Fig. \ref{fig:FSI-tri}. The kernel for this process involves the $KDD$ triangle diagram, as shown in the figure, and may be written as an effective interaction given by

\begin{eqnarray}
V_{eff}(q,q') &=& \int \frac{d^3Q}{(2\pi)^3} F_{3P0}(\bm{q-Q},\bm{Q}) F_{3P0}(\bm{q-Q}, \bm{q'-q+Q}) \cdot \nonumber \\
&& \qquad  F(\bm{Q},\bm{-q}) F(\bm{q'-q+Q},\bm{-q'}) \cdot \lambda_{D\Lambda_c:D\Lambda_c} \cdot \nonumber \\
&& \qquad [E - E_K(\bm{q-Q}) - E_D(\bm{Q}) + i\epsilon]^{-1} \nonumber \\
&& \qquad [E -  E_K(\bm{q-Q}) - E_D(\bm{q'-q+Q}) + i\epsilon]^{-1}.
\end{eqnarray}
Here we assume that the 3p0 decay model describes the $D_sKD$ vertices and a contact FSI interaction, $F \lambda F$, as employed in Eq. \ref{eq:t}. The strengths of the strong decay vertices are 

\begin{equation}
F_{3P0} \sim \gamma/\sqrt{\beta},
\end{equation}
where $\gamma$ is the 3p0 decay constant. 
Recall that $\beta \sim \Lambda_{QCD}$ sets the scales for the hadronic wavefunctions, the decay couplings, and the final state interactions.

Performing the integral gives the approximate scaling result
\begin{equation}
V_{eff} \sim \frac{\gamma}{\sqrt{\beta}} \frac{\gamma}{\sqrt{\beta}} \lambda_{D\Lambda_c:D\Lambda_c} \cdot \frac{\mu_{KD}^2}{\beta} f((E-m_K-m_D)\mu_{KD}/\beta^2)
\end{equation}
or $V_{eff} \sim \gamma^2 A\,  \mu_{KD}^2 f /\beta^2$ (referring to Tables \ref{tab:1} or \ref{tab:2}). Under normal conditions the function $f$ is order one and $V_{eff} \sim \gamma^2 A$.
Since the 3P0 coupling is $\gamma \approx 0.4$\cite{Barnes:2005pb}, we conclude that the effective interaction is quite weak with respect to those considered above. This conclusion is valid \textit{absent} kinematical enhancements that make $f$ large, that of course can occur in triangle diagrams. In our case, this does not happen because the system energy is set by the $\Lambda_b$ mass, which is very far removed from $m_K + m_D$.

\appendix

\end{document}

%% file: fits.tex
		&		&	Case 1	&	Case 2a	&	Case 2b	&	Case 3	&	Case 4	&	Case 5	\\
		&		&	Fig.~\ref{fig:fit1}	&	Fig.~\ref{fig:fit2a}	&	Fig.~\ref{fig:fit2b}	&	Fig.~\ref{fig:fit3}	&	Fig.~\ref{fig:fit4}	&	Fig.~\ref{fig:fit5}	\\
	\hline															
	$M_{D_s}$ (GeV)															
		&$	1/2^-,3/2^-	$&	2.112	&	2.112	&	2.112	&	2.112	&	2.112	&	2.112	\\
		&$	1/2^+	$&		&		&		&	2.112	&	2.920	&	2.112	\\
	\hline															
	Contact terms															
		&$	B	$&$	4.0	$&$	4.0	$&$	6.0	$&$	4.0	$&$	4.0	$&$	4.0	$\\
		&$	C_a	$&$		$&$	-14.8	$&$	-14.0	$&$	-14.8	$&$	-15.2	$&$	-14.8	$\\
		&$	C_b	$&$		$&$	-8.0	$&$	-9.8	$&$	-8.0	$&$	-7.4	$&$	-8.0	$\\
		&$	D	$&$	1.0	$&$	0.6	$&$	0.7	$&$	0.6	$&$	0.6	$&$	0.6	$\\
		&$	E	$&$	1.0	$&$	1.0	$&$	1.0	$&$	1.0	$&$	1.0	$&$	1.0	$\\
		&$	F_a	$&$		$&$		$&$		$&$		$&$		$&$	-12.0	$\\
		&$	G_a	$&$		$&$		$&$		$&$	0.3	$&$	0.3	$&$	0.3	$\\
		&$	G_b	$&$		$&$		$&$		$&$	0.2	$&$	0.0	$&$	0.2	$\\
	\hline															
	Production															
		&$	g_1	$&$	0.0710	$&$	0.0560	$&$	-0.0083	$&$	0.0198	$&$	0.0217	$&$	0.0382	$\\
		&$	g_2	$&$	-0.1899	$&$	-0.2181	$&$	-0.3302	$&$	0.1718	$&$	0.1893	$&$	-0.2137	$\\
		&$	g_3	$&$	0.1015	$&$	0.0309	$&$	0.0342	$&$	-0.1239	$&$	0.0305	$&$	0.0320	$\\
		&$	g_4	$&$		$&$		$&$		$&$	0.4358	$&$	0.0549	$&$	-0.0296	$\\
		&$	g_5	$&$		$&$		$&$		$&$	-0.1009	$&$	0.0039	$&$	-0.0066	$\\
		&$	g_6	$&$		$&$		$&$		$&$	-0.1640	$&$	-0.0259	$&$	0.0369	$\\
	\hline															
	Background															
		&	Re $b_1$	&$	0.001653	$&$	0.000739	$&$	0.000707	$&$	-0.000283	$&$	0.000505	$&$	0.000641	$\\
		&	Im $b_1$	&$	-0.001378	$&$	-0.000892	$&$	-0.000839	$&$	0.000670	$&$	-0.000835	$&$	-0.000883	$\\
		&	Re $b_2$	&$	0.001515	$&$	0.001207	$&$	0.000750	$&$	-0.000655	$&$	0.001040	$&$	0.001176	$\\
		&	Im $b_2$	&$	-0.000004	$&$	-0.000365	$&$	0.000827	$&$	0.000446	$&$	-0.000629	$&$	-0.000402	$\\
		&	Re $b_3$	&$		$&$		$&$		$&$	-0.000841	$&$		$&$		$\\
		&	Im $b_3$	&$		$&$		$&$		$&$	-0.000184	$&$		$&$		$\\
		&	Re $b_4$	&$		$&$		$&$		$&$	0.000691	$&$		$&$		$\\
		&	Im $b_4$	&$		$&$		$&$		$&$	-0.001093	$&$		$&$		$\\
	\hline															
	N. params	&		&	10	&	12	&	12	&	21	&	17	&	18	\\
	\hline
	$\chi^2$/dof & as fit        & 4.76 & 2.83 & 3.33 & 2.30 & 2.55 & 2.69 \\
	             & full data set & 4.63 &  2.77 & 4.00 & 2.78 & 2.41 & 2.51 \\
	             & (4.25 - 4.55) & 3.63 & 2.29 & 2.88 & 1.69 & 1.96 & 2.12 \\  
	\hline															